\documentclass[amsfonts,amsmath,prd,preprint,nofootinbib]{revtex4}
\usepackage{epsfig,bbm,cancel,ulem}
\usepackage[breaklinks=true]{hyperref}
\usepackage[utf8]{inputenc}
\usepackage[english]{babel}
 \usepackage{graphicx}
\usepackage{xcolor}
\usepackage{subfigure} 
\usepackage{CJKutf8} 
\usepackage{appendix}
\usepackage[papersize={8.5in,11in}]{geometry}

\begin{document}
\begin{CJK}{UTF8}{gbsn}
\title{Effects of a global monopole on thermodynamic phase transition of the charged AdS black hole}
\author{Zhi Luo}
\email{ZhiLuo@cqu.edu.cn}
\author{Hao Yu}
\email{yuhaocd@cqu.edu.cn}
\author{Jin Li}
\email{cqstarv@hotmail.com}
\affiliation{Department of Physics, Chongqing University, Chongqing, 401331, China. }
\def\changenote#1{\footnote{\bf #1}}

\begin{abstract}
In this paper, we study the dynamical properties of thermodynamic phase transition (PT) for the charged AdS black hole (BH) with a global monopole via the Gibbs free energy landscape and reveal the effects of a global monopole on the kinetics of the AdS BH thermodynamic PT. First, we briefly review the thermodynamics of the charged AdS BH with a global monopole. Then, we introduce the Gibbs free energy landscape to study the thermodynamic stability of the BH state. Because of thermal fluctuations, the small black hole (SBH) state can transit to the large black hole (LBH) state, and vice versa. We use the Fokker-Planck equation with the reflecting boundary condition to study the probability evolution of the BH state with and without a global monopole separately. We find that for both the SBH and LBH states, the global monopole could slow down the evolution of the BH state. In addition, we obtain the relationship between the first passage time and the monopole parameter $\eta$. The result shows that as the monopole parameter $\eta$ increases, the mean first passage time will be longer for both the SBH and LBH states. 

\end{abstract}

\maketitle
\thispagestyle{empty}
\setcounter{page}{1}

\section{Introduction}

The images of Sagittarius A* in our galaxy provide new opportunities for testing the validity of gravity theories and proving the existence of the thermodynamic phase transition (PT) for the black hole (BH)~\cite{EventHorizonTelescope:2022xnr,EventHorizonTelescope:2022vjs,EventHorizonTelescope:2022wok,EventHorizonTelescope:2022exc,EventHorizonTelescope:2022urf,EventHorizonTelescope:2022xqj}. BH thermodynamics has been researched since the 1970s~\cite{Bekenstein:1972tm,Bekenstein:1973ur,Bardeen:1973gs,Hawking:1975vcx}, and the Hawking-Page PT was first found in 1983~\cite{Hawking:1982dh}. In the new century, due to the AdS/CFT correspondence~\cite{Maldacena:1997re,Witten:1998qj,Gubser:1998bc}, the thermodynamics of AdS BHs has been a widely concerned topic in physics~\cite{Kastor:2009wy,Dolan:2010ha,Cvetic:2010jb,Chamblin:1999tk,Chamblin:1999hg,Caldarelli:1999xj}. In the extended phase space, the cosmological constant can be treated as the pressure of the BH thermal system, i.e., $P=-\Lambda / 8 \pi$~\cite{Maldacena:1997re,Dolan:2010ha,Cvetic:2010jb}. The research shows that there could exist some special PTs associated with AdS BHs~\cite{Kubiznak:2012wp,Altamirano:2013uqa,Hennigar:2016xwd,Hendi:2017fxp,Gunasekaran:2012dq,Altamirano:2014tva,Kubiznak:2015bya,Wei:2014hba,Frassino:2014pha,Hennigar:2015esa,Altamirano:2013ane,Sherkatghanad:2014hda,Hennigar:2015wxa}. In the extended phase space, the $P-V$ criticality of charged AdS BHs was studied by Kubiznak et al., and the results showed that there exists a van der Waals (vdW)-like PT and the charged AdS BH system is similar to the liquid-gas system~\cite{Kubiznak:2012wp}. At present, more different types of AdS BH PTs have been studied, such as the reentrant PT (RPT) and the triple point PT~\cite{Altamirano:2013ane,Altamirano:2013uqa,Wei:2014hba}. For an exhaustive overview of the thermodynamic PT of AdS BHs, one can refer to Ref.~\cite{Kubiznak:2016qmn}.    

For AdS BHs, an interesting question is how to understand the underlying kinetics of BH PTs. Recently, Li et al. used the Gibbs free landscape to study the dynamical properties of the Hawking-Page PT for Schwarzschild AdS BHs and the vdW-like PT for Reissner-Nordstr\"om AdS BHs~\cite{Li:2020khm,Li:2020nsy}. In the case of the Hawking-Page PT, the thermal AdS space phase can transit to the large Schwarzschild AdS BH phase due to thermal fluctuations, and the probability evolution can be studied using the Fokker-Planck equation with the reflecting boundary condition~\cite{Li:2020khm}. For the vdW-like PT, Li et al. calculated the Fokker-Planck equation with the reflecting and absorbing boundary conditions to study the switching dynamics of the BH state~\cite{Li:2020nsy}.     

Following the works of Li et al., this method was generalized to rotating BHs and beyond-GR BHs~\cite{Yang:2021ljn,Wei:2021bwy,Li:2021vdp,Lan:2021crt,Wei:2020rcd}. In the case of the Kerr-AdS BH, Yang et al. found that the BH microstructure can be understood via the dynamical properties of BH PTs~\cite{Yang:2021ljn}. For the RN-AdS BH surrounded by quintessence, Lan et al. studied the impact of quintessence dark energy on the dynamical properties of BH PTs~\cite{Lan:2021crt}. The results show that the mean first passage time is not a simple monotonic relationship with respect to the state parameter of quintessence dark energy, and there exists a local maximum value.   

Inspired by these pioneering works, we extend the works of Li et al. to the charged AdS BH with a global monopole. During the PT in the early universe, a global monopole can be generated from the breaking of a global $O(3)$ symmetry to $U(1)$~\cite{Vilenkin:1986hg}. Barriola and Vilenkin studied the BH with a global monopole and found that the monopole can generate a deficit solid angle in the space~\cite{Barriola:1989hx}. Recently, Soroushfar and Upadhyay used geometrical thermodynamics to study the thermodynamic PT of the charged AdS BH with a global monopole and found that the monopole parameter $\eta$ can affect the thermodynamic PT~\cite{Soroushfar:2020wch}. In this paper, we inspect the effects of a global monopole on the thermodynamic PT of the charged AdS BH by  studying the probability evolution and the first passage process of the BH state.                

The paper is organized as follows: In Sec. \ref{sec2}, the thermodynamics of the charged AdS BH with a global monopole is briefly reviewed. In Sec. \ref{sec3}, we study the thermodynamic PT of the charged AdS BH with a global monopole, and we focus on the probability evolution and the first passage process of the BH state. The conclusions are given in Sec. \ref{sec5}. In this paper, we employ the units $G_{\mathrm{N}}=\hbar=\kappa_{\mathrm{b}}=c=1$.

\section{Thermodynamics of the charged AdS BH with a global monopole} \label{sec2}
In this section, we present a brief overview of the charged AdS BH with a global monopole and its thermodynamics. The Lagrangian density of the charged AdS BH with a global monopole is~\cite{Barriola:1989hx}
\begin{equation}
	\mathcal{L}=R-2 \Lambda+\frac{1}{2} \partial_{\mu} \phi^{a} \partial^{\mu} \phi^{* a}-\frac{\gamma}{4}\left(\phi^{a} \phi^{* a}-\eta_{0}^{2}\right)^{2},
\end{equation}
where $\gamma$ is a constant and $\eta_{0}$ is the energy scale of symmetry breaking. The parameter $\phi^{a}$ represent a triplet of the scalar field as follows:
\begin{equation}
	\phi^{a}=\eta_{0} h(\tilde{r}) \frac{\tilde{x}^{a}}{\tilde{r}},
\end{equation}
where $\tilde{x}^{a} \tilde{x}^{a}=\tilde{r}^{2}$. The metric of the charged AdS BH with a global monopole is given by~\cite{Barriola:1989hx}
\begin{equation}
	\label{metric}
	\mathrm{d} s^{2}=-f(\tilde{r}) \mathrm{d} \tilde{t}^{2}+\frac{1}{f(\tilde{r})} \mathrm{d} \tilde{r}^{2}+\tilde{r}^{2}\left(\mathrm{d} \theta^{2}+\sin ^{2} \theta \mathrm{d} \phi^{2}\right),
\end{equation}
where the metric function $f(\tilde{r})$ is
\begin{equation}
	f(\tilde{r})=1-8 \pi \eta_{0}^{2}-\frac{2 \tilde{m}}{\tilde{r}}+\frac{\tilde{q}^{2}}{\tilde{r}^{2}}+\frac{\tilde{r}^{2}}{l^{2}}.
\end{equation}
Here, $\tilde{m}$ is the BH mass, $\tilde{q}$ is the BH charge, and $l(=\sqrt{-\frac{\Lambda}{3}})$ is the AdS curvature radius. For convenience, we introduce the following coordinate transformation  
\begin{equation}
	\tilde{t}=\left(1-8 \pi \eta_{0}^{2}\right)^{-1 / 2} t,~~~~ \tilde{r}=\left(1-8 \pi \eta_{0}^{2}\right)^{1 / 2} r,
\end{equation}
and new parameters
\begin{equation}
	m=\left(1-8 \pi \eta_{0}^{2}\right)^{-3 / 2} \tilde{m}, ~~q=\left(1-8 \pi \eta_{0}^{2}\right)^{-1} \tilde{q}, ~~\eta^{2}=8 \pi \eta_{0}^{2}.
\end{equation}
Therefore, the metric (\ref{metric}) can be rewritten as
\begin{equation}\label{ef}
	\mathrm{d} s^{2}=-f(r) \mathrm{d} t^{2}+\frac{1}{f(r)} \mathrm{d} r^{2}+\left(1-\eta^{2}\right) r^{2}\left(\mathrm{~d} \theta^{2}+\sin ^{2} \theta \mathrm{d} \phi^{2}\right),
\end{equation}
where
\begin{equation}
	f(r)=1-\frac{2 m}{r}+\frac{q^{2}}{r^{2}}+\frac{r^{2}}{l^{2}}.
\end{equation}
It is noteworthy that the non-vanishing monopole parameter $\eta$ will result in a solid angle deficit in Eq. (\ref{ef}).  Therefore, the ADM mass and the electric charge of the BH are modulated by the monopole parameter $\eta$:
\begin{equation}
	M=\left(1-\eta^{2}\right) m, \quad~~~~~~~ Q=\left(1-\eta^{2}\right) q.
\end{equation}

The event horizon $r_{h}$ can be obtained by solving $f\left(r\right)=0$, and then the ADM mass $M$ can be rewritten as
\begin{equation}
	M=\frac{Q^{2}}{2 r_{h}-2 \eta^{2} r_{h}}-\frac{1}{2}\left(\eta^{2}-1\right) r_{h}\left(1+ l^{-2} r_{h}^{2}\right).
\end{equation}
The temperature and entropy of the BH can be expressed as 
\begin{equation}
	T=\frac{1}{4 \pi r_{h}}\left(1+\frac{3 r_{h}^{2}}{l^{2}}-\frac{Q^{2}}{\left(1-\eta^{2}\right)^{2} r_{h}^{2}}\right),~~~~~~S=\pi\left(1-\eta^{2}\right) r_{h}^{2}.
\end{equation}
Furthermore, in the extended phase space (i.e., $P=3/8 \pi l^{2}$), one can obtain
\begin{equation}
	\mathrm{d} M= T\mathrm{d}S+\Phi\mathrm{d}Q+V\mathrm{d}P,
\end{equation}
where
\begin{equation}
	\begin{aligned}
		\Phi=\frac{q}{r_{h}}, ~~~~~~~~~V=\frac{4}{3} \pi\left(1-\eta^{2}\right) r_{h}^{3}.
	\end{aligned}
\end{equation}
The Smarr relation still holds as follows:
\begin{equation}\label{esmarr}
	M=2(T S-V P)+\Phi Q.
\end{equation}
Note that $M$ in Eq. (\ref{esmarr}) is now regarded as the enthalpy of the BH.

Therefore, for the charged AdS BH with a global monopole, the equation of state is given as
\begin{equation}\label{ept}
	P=\frac{T}{2 r_{h}}-\frac{1}{8 \pi r_{h}{ }^{2}}+\frac{Q^{2}}{8 \pi\left(1-\eta^{2}\right) r_{h}{ }^{4}}.
\end{equation}
By solving the equations $\left(\partial_{r_{\mathrm{h}}} P\right)_{T}=0=$ $\left(\partial_{r_{\mathrm{h}}, r_{\mathrm{h}}} P\right)_{T}$, one can obtain the critical point
\begin{equation}
	P_{c}=\frac{\left(1-\eta^{2}\right)^{2}}{96 \pi Q^{2}},~~r_{c}=\frac{\sqrt{6} Q}{\sqrt{1-2 \eta^{2}+\eta^{4}}}, ~~T_{c}=\frac{\sqrt{\left(1-\eta^{2}\right)^{2}}}{3 \sqrt{6} \pi Q}.
\end{equation}
In this paper, we set $Q=1$ for convenience. 
\begin{figure}
	\centering
	
	\subfigure[$P=1.6P_{c}$.]{
		\begin{minipage}[t]{0.5\linewidth}
			\centering
			\includegraphics[width=2.5in]{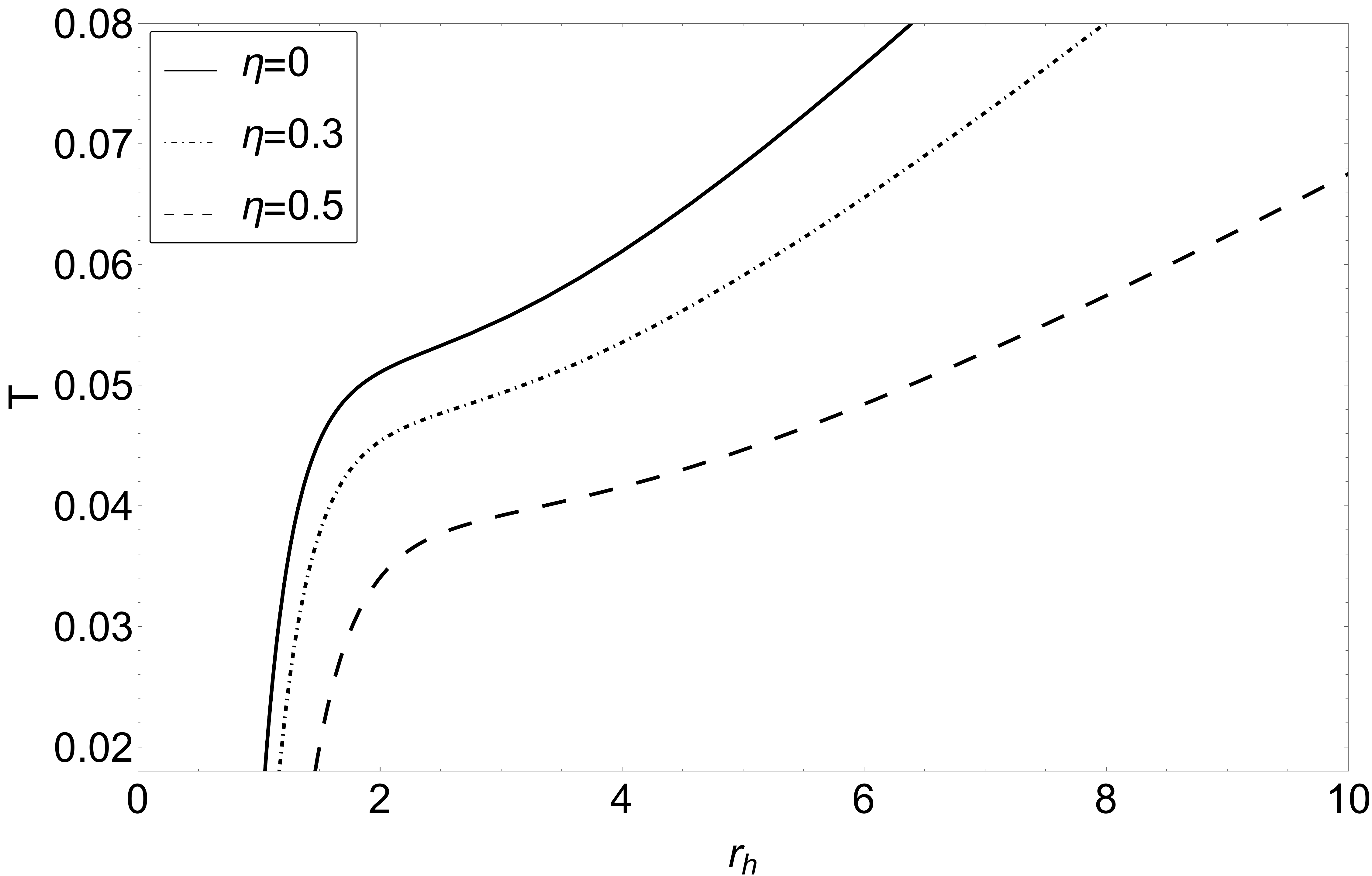}
		\end{minipage}%
	}%
	\subfigure[$P=P_{c}$.]{
	\begin{minipage}[t]{0.5\linewidth}
		\centering
		\includegraphics[width=2.5in]{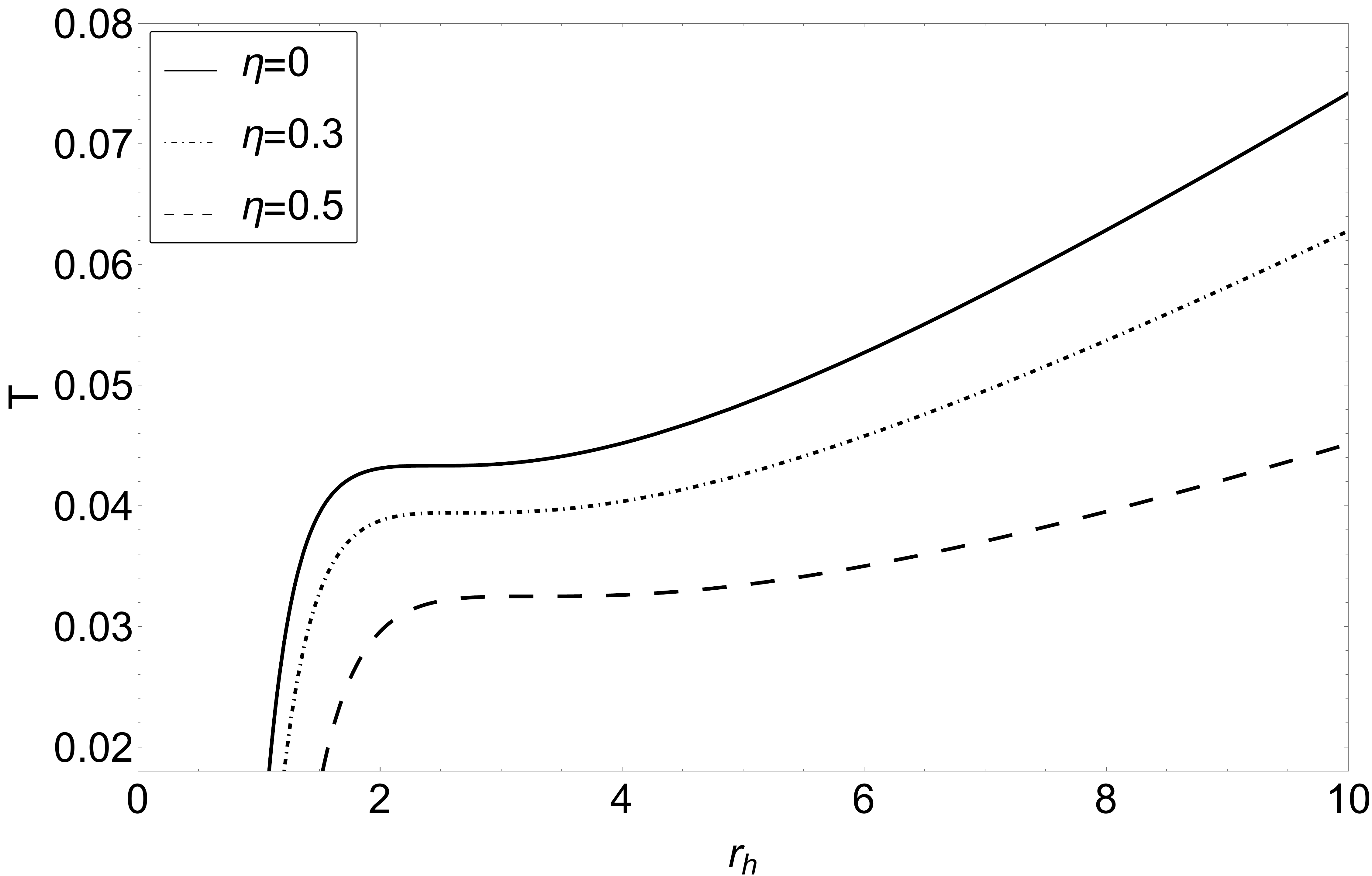}
	\end{minipage}%
}
\\
	\subfigure[$P=0.4P_{c}$.]{
		\begin{minipage}[t]{0.5\linewidth}
			\centering
			\includegraphics[width=2.5in]{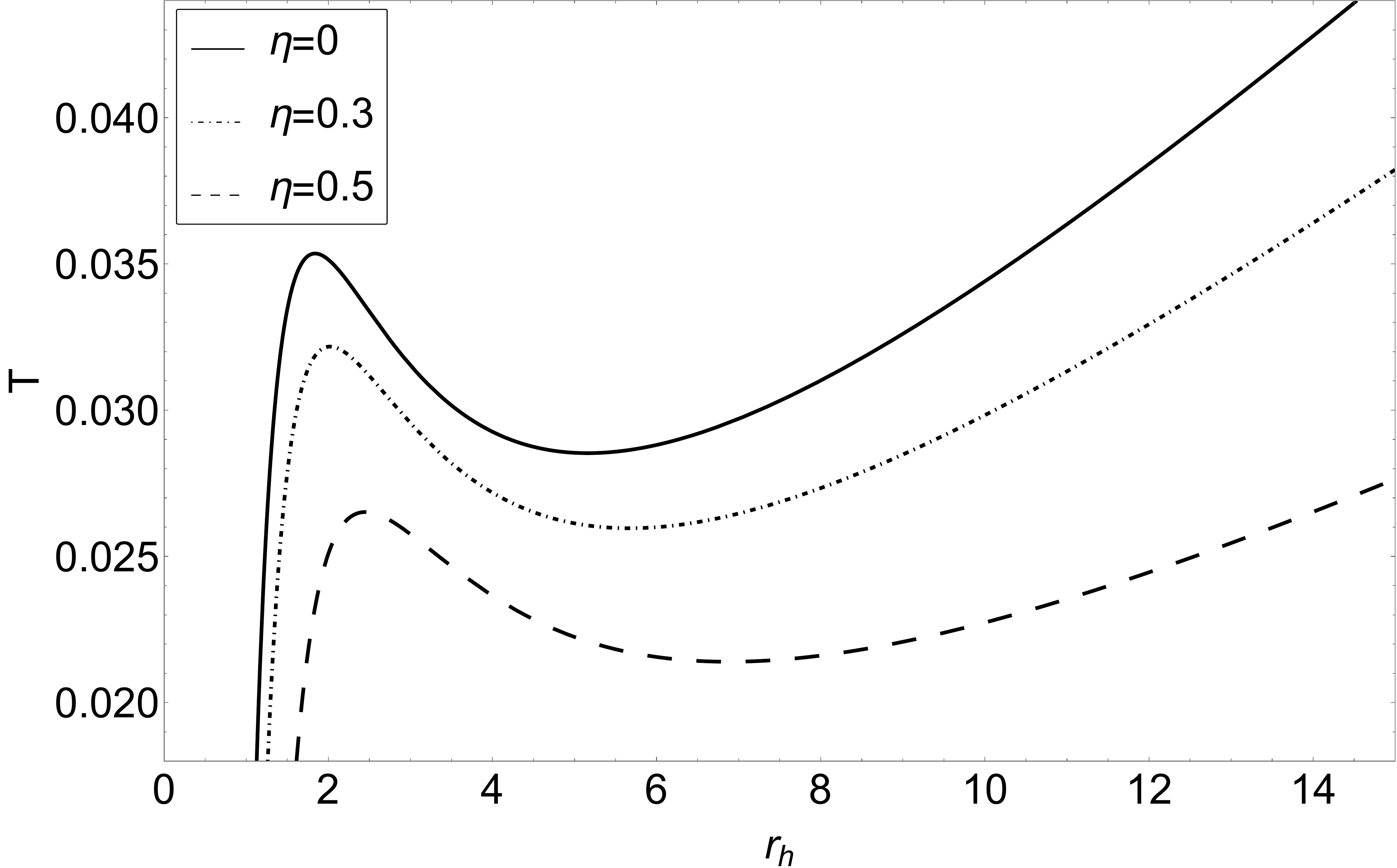}
		\end{minipage}\label{figc}
	}%
	\subfigure[$P=0.4P_{c}$ and $\eta=0.5$.]{
	\begin{minipage}[t]{0.5\linewidth}
		\centering
		\includegraphics[width=2.5in]{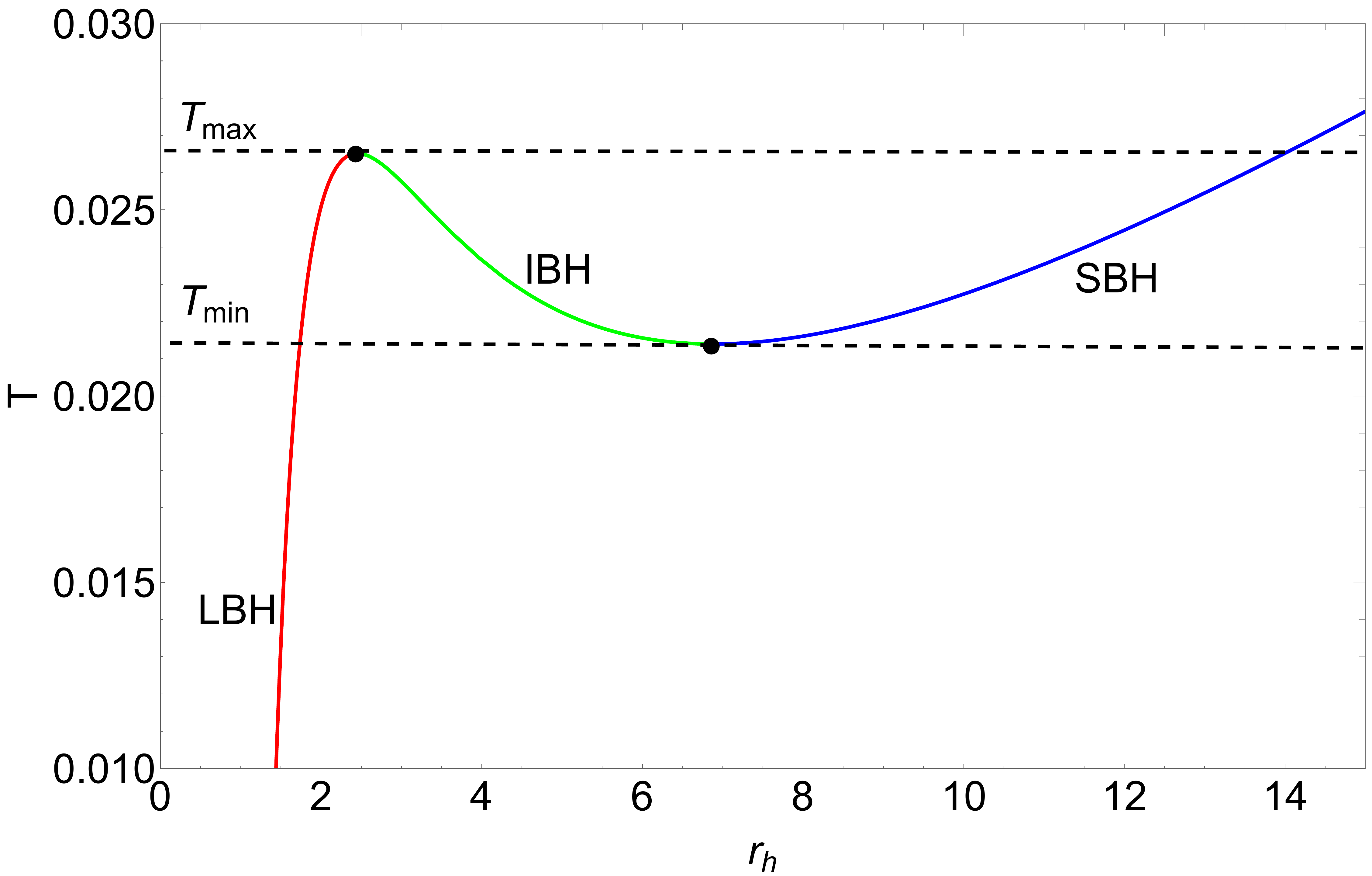}
	\end{minipage}\label{figd}
}
	\centering
	\caption{The temperature $T$ as a function of the BH radius $r_{h}$. The pressures in the first three panels are selected as (a) $P=1.6P_c$; (b)  $P=P_c$; (c) $P=0.4P_c$. There are three sets of the monopole parameters ($\eta=0,~0.3,~0.5$) for comparison. The last panel corresponds to the dashed line in the figure (c). }\label{fig:tr} 
\end{figure}

	As shown in Fig.~\ref{fig:tr}, the evolution of the BH temperature over the BH radius $r_{h}$ is closely retated to the value of the pressure $P$. When $P\geq P_{c}$, $T$ is a monotonic function of the radius $r_{h}$, while when $P<P_{c}$ there is a local minimum temperature and a local maximum temperature. The local minimum and maximum of the BH temperature $T$ are determined by 
\begin{equation}\label{etr}
	\frac{\partial T}{\partial r_{h}}=0,
\end{equation}
which yields 
\begin{equation}\label{erpm}
	r_{\min / \max }=\frac{1}{4} \sqrt{\frac{1-\eta^{2} \pm \sqrt{1-96 P \pi Q^{2}-2 \eta^{2}+\eta^{4}}}{P \pi-P \pi \eta^{2}}}.
\end{equation}
The positive (negative) sign in Eq. (\ref{erpm}) corresponds to the BH radius $r_{\min}$ ($r_{\max}$). By substituting Eq. (\ref{erpm}) into Eq. (\ref{ept}), we get the local minimum and maximum of the BH temperature:
\begin{equation}\label{etm}
	T_{\min / \max }=\frac{-64 P \pi Q^{2} \pm 2\left(-1+\eta^{2}\right)\left(1-\eta^{2}+\sqrt{-96 P \pi Q^{2}+\left(-1+\eta^{2}\right)^{2}}\right)}{P \sqrt{\pi}\left(-1+\eta^{2}\right)^{2}\left(\frac{1-\eta^{2}+\sqrt{-96 P \pi Q^{2}+\left(-1+\eta^{2}\right)^{2}}}{P-P \eta^{2}}\right)^{3 / 2}},
\end{equation}
where the positive (negative) sign in Eq. (\ref{etm}) corresponds to the local minimum $T_{\min}$ (the local maximum $T_{\max}$). When $T_{\min }<T<T_{\max}$, there are three branchs: small BH (SBH) state, intermediate BH (IBH) state, and large BH (LBH) state (see Figs.~\ref{figc} and~\ref{figd}). It is worth noting that the IBH state is not stable, and there exists a first-order SBH/LBH PT.       

\section{Dynamical properties of thermodynamic phase transition} \label{sec3}
\subsection{Gibbs free energy}
\begin{figure}[]
	\centering

	\subfigure[\label{fig:gr1}]{
		\begin{minipage}[t]{0.5\linewidth}
			\centering
			\includegraphics[width=2.5in]{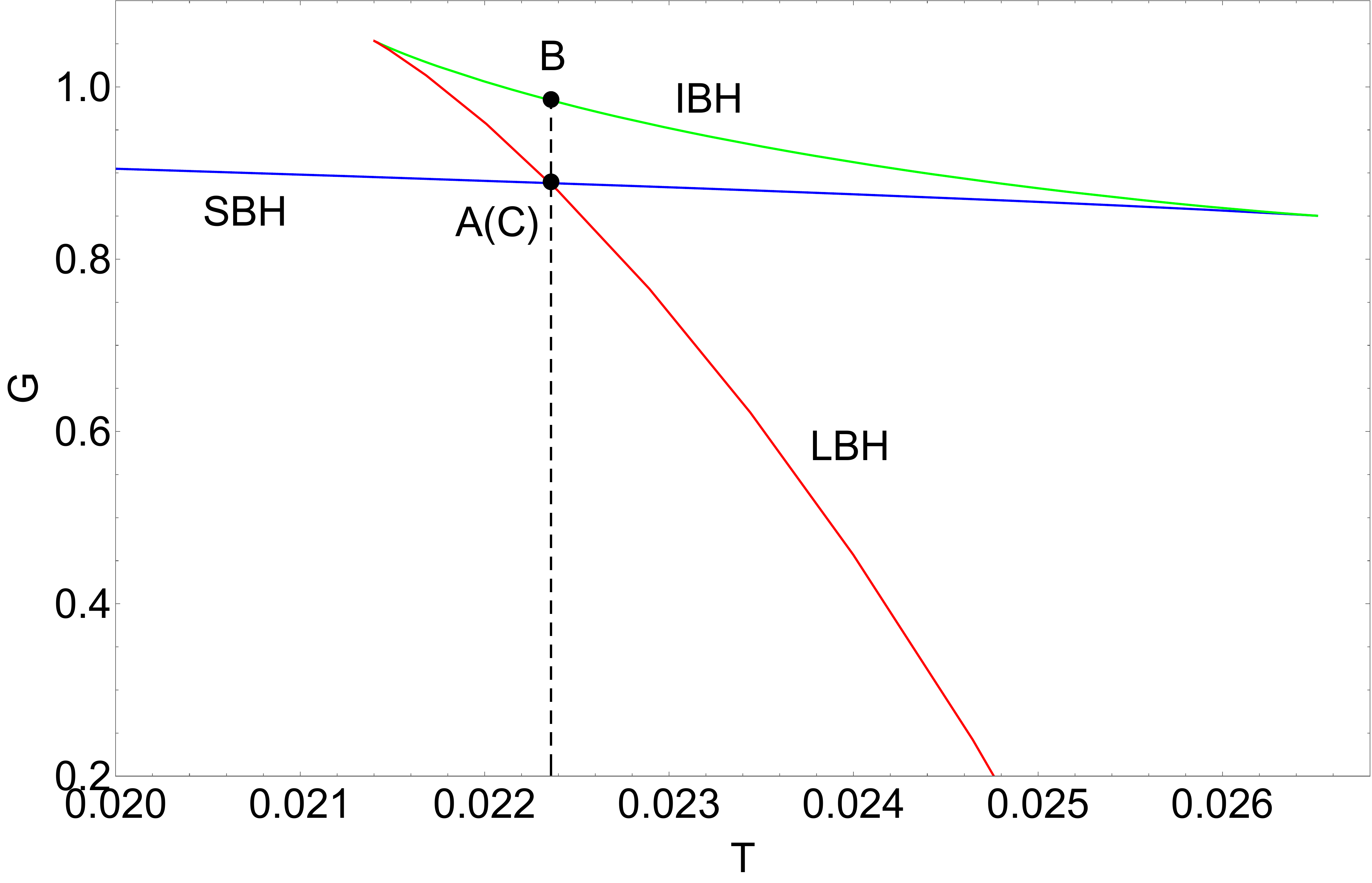}
		\end{minipage}%
	}%
	\subfigure[\label{fig:gr2}]{
		\begin{minipage}[t]{0.5\linewidth}
			\centering
			\includegraphics[width=2.5in]{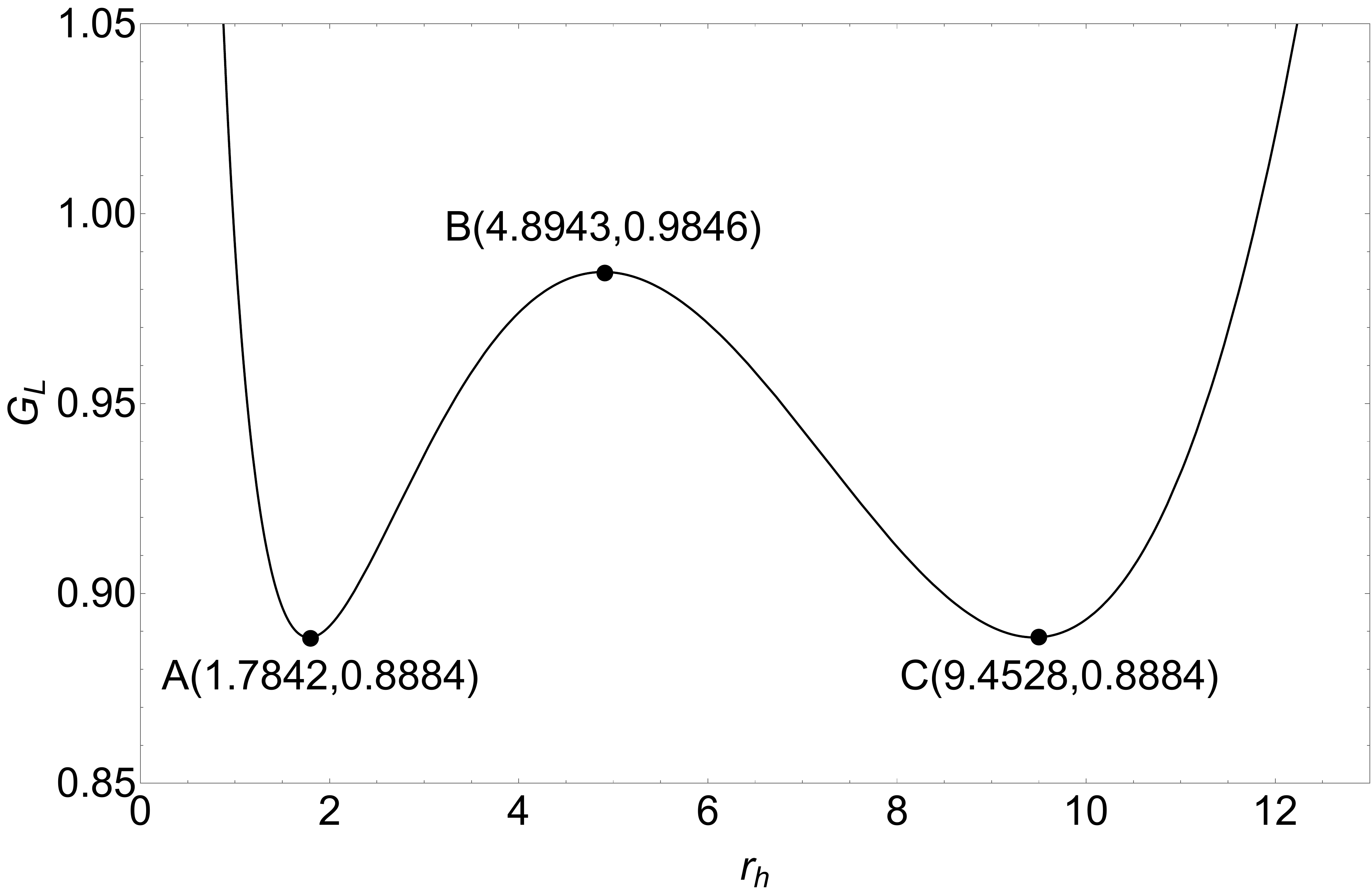}
		\end{minipage}%
	}%

	\caption{Gibbs free energy of the charged AdS BH with a global monopole. For all plots, we set  $\eta=0.5$ and $P=0.4P_{c}$. The left panel is the $G-T$ diagram, and the right panel is the $G_{L}-r_{h}$ diagram with $T_{E}=0.022355$. }
	\label{fig:gr}
\end{figure}

In the study of the BH thermodynamics, Gibbs free energy is usually used to investigate the thermodynamic PT of BHs~\cite{Kubiznak:2012wp,Altamirano:2013uqa,Hendi:2017fxp,Gunasekaran:2012dq,Altamirano:2014tva,Kubiznak:2015bya,Wei:2014hba,Frassino:2014pha,Hennigar:2015esa,Altamirano:2013ane,Sherkatghanad:2014hda,Hennigar:2015wxa}. For the charged AdS BH with a global monopole, the Gibbs free energy is given by
\begin{equation}
	G=H-T S=\frac{1}{6} r_{h}\left(3+8 P \pi r_{h}^{2}-6 \pi r_{h} T\right)\left(1-\eta^{2}\right)+\frac{Q^{2}}{2 r_{h}\left(1-\eta^{2}\right)}.
\end{equation}
The $G-T$ diagram of the charged AdS BH with a global monopole is plotted in Fig.~\ref{fig:gr1} with $\eta=0.5$ and $P=0.4P_{c}$, from which we can see there exist three branches of the BH state and a swallowtail behavior. The blue, green, and red lines represent the SBH, IBH, and LBH branches, respectively. The temperature corresponding to the intersection (see the point A(C) in Fig.~\ref{fig:gr1}) of the SBH branch and LBH branch is about $T_{A}=T_{C}=0.022355$. The point $B$ is in a state of the IBH, and the point $A$ and $C$ are in states of the SBH and LBH, respectively.

Recently, the concept of the Gibbs free energy landscape is introduced to study the dynamical properties of the thermodynamic PH of BHs~\cite{Li:2020khm,Li:2020nsy}. In the Gibbs free energy landscape, the Hawking temperature of the BH is replaced by the temperature of the ensemble. Therefore, for the charged AdS BH with a global monopole, the Gibbs free energy $G_{L}$ can be expressed as 
\begin{equation}\label{egr1}
	G_{L}=H-T_{E} S=-\frac{1}{2} r\left(-1+\eta^{2}\right)-\frac{4}{3} P \pi r^{3}\left(-1+\eta^{2}\right)+\pi r^{2} T_{E} \left(-1+\eta^{2}\right)+\frac{Q^{2}}{2 r-2 r \eta^{2}},
\end{equation}
where $T_{E}$ is the temperature of the ensemble. The $G_L-r_h$ diagram is plotted in Fig.~\ref{fig:gr2} with $\eta=0.5$, $P=0.4P_{c}$, and $T_{E}=0.022355$. As can be seen, the $G_{L}-r_{h}$ diagram has a double-well shape and there exist two equivalent local minimums and a local maximum. Since the SBH and LBH correspond to the local minimums of the Gibbs free energy, they are stable. As for the IBH which corresponds to the local maximum of the Gibbs free energy, it is unstable.

\subsection{Fokker-Planck equation and probability evolution}\label{secfp}
Recently, the Fokker-Planck equation has been proposed to study the probability evolution of the BH phase transition, which is given by~\cite {Li:2020khm,Li:2020nsy,Bryngelson1989,Zwanzig,Lee2002,Lee2003,Wang2015}
\begin{equation}\label{erhot}
	\frac{\partial \rho(r, t)}{\partial t}=D \frac{\partial}{\partial r}\left\{e^{-\beta G(r)} \frac{\partial}{\partial r}\left[e^{\beta G(r)} \rho(r, t)\right]\right\},
\end{equation}
where $\rho(r, t)$ is the probability distribution, $\beta=\frac{1}{k_{\mathrm{B}} T}$ presents the inverse temperature, and $D=\frac{k_{\mathrm{B}} T}{\zeta}$ is the diffusion coefficient. The Boltzman constant $k_{\mathrm{B}}$ and the dissipation coefficient $\zeta$ as usual are set to $1$.

Mathematically, one can solve the Fokker-Planck equation by choosing specific boundary conditions. To study the probability evolution of the BH phase transition, we use the reflecting boundary condition and the absorbing boundary condition at $r=r_{0}$ to solve the Fokker-Planck equation, which can be given as 
\begin{equation}
	\begin{aligned}
		\beta G_{\mathrm{L}}^{\prime}\left(T, P, r_{0}\right) \rho\left(r_{0}, t\right)+\rho^{\prime}\left(r_{0}, t\right) =0, \\
		\rho\left(r_{0}, t\right)=0.
	\end{aligned}
\end{equation}
Here, the prime indicates the time derivative.  

We first consider the reflecting boundary condition, for which there are two boundaries at $r=0$ and $r=\infty$, respectively. The Gaussian wave packet is selected as the initial wave packet, which is given by
\begin{equation}
\rho(r, 0)=\frac{1}{\sqrt{\pi} a} e^{-\frac{\left(r-r_{s/i/l}\right)^{2}}{a^{2}}},
\end{equation}
where the parameter $a$ is set to $1$ and $r_{s/i/l}$ is the Gaussian wave packet located at the SBH/IBH/LBH radius.

\begin{figure}[h]
	\centering

	\subfigure[$\eta=0,~T_{E}=0.029806$\label{ptr1}]{
		\begin{minipage}[t]{0.33\linewidth}
			\centering
			\includegraphics[width=2.0in]{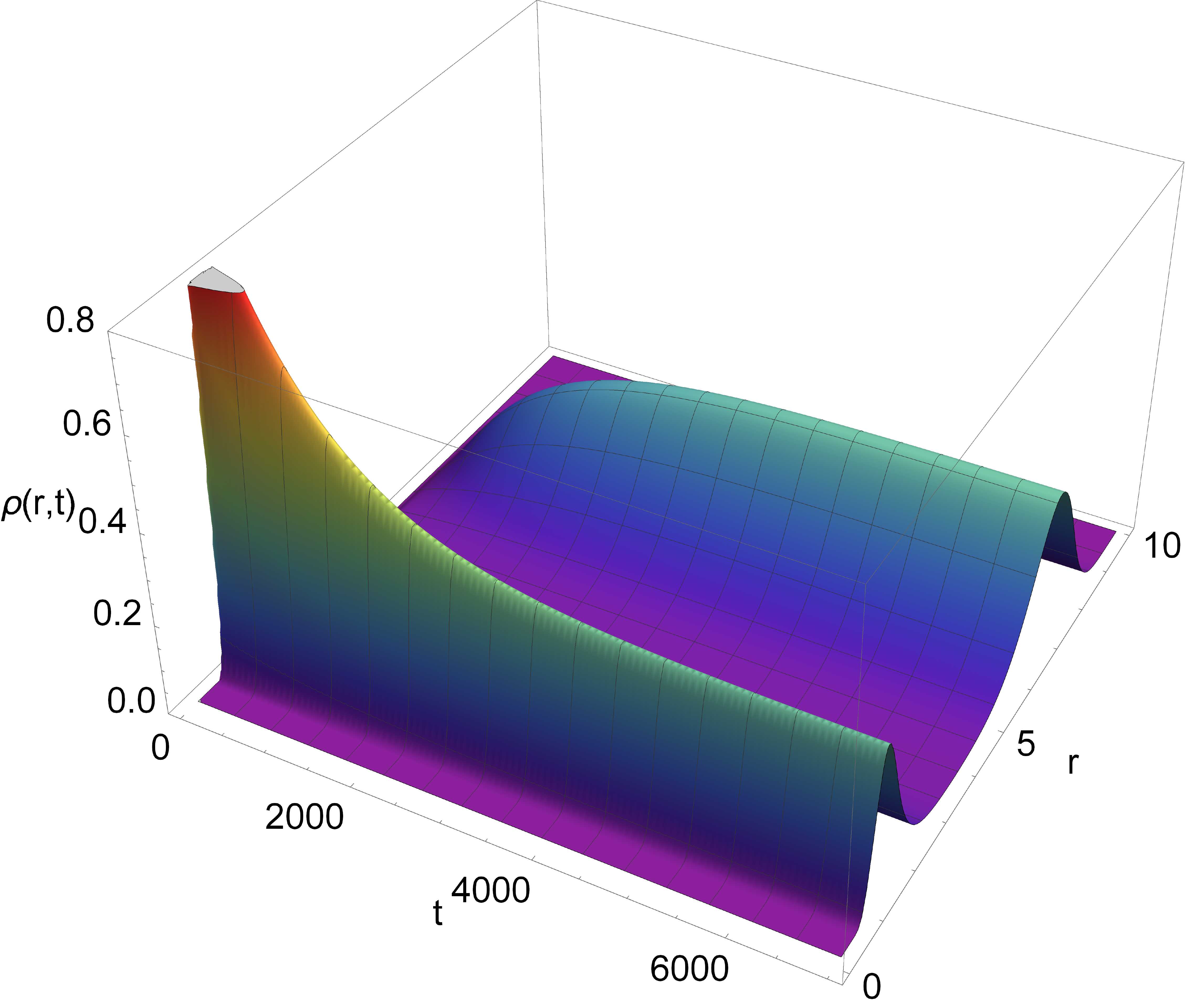}
		\end{minipage}%
	}%
	\subfigure[$\eta=0.3,~T_{E}=0.027124$\label{ptr2}]{
		\begin{minipage}[t]{0.33\linewidth}
			\centering
			\includegraphics[width=2.0in]{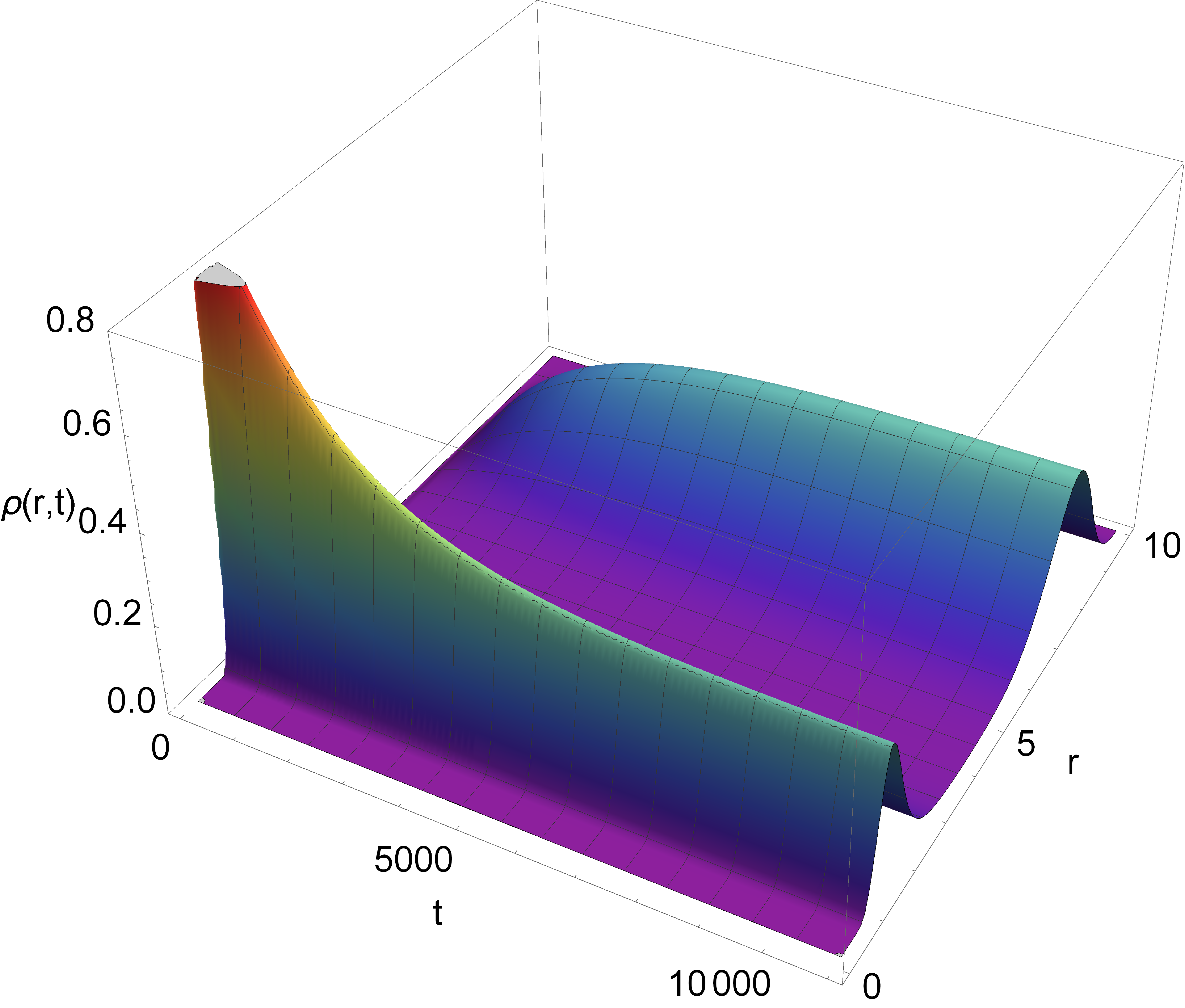}
		\end{minipage}%
	}%
	\subfigure[$\eta=0.5,~T_{E}=0.022355$\label{ptr3}]{
		\begin{minipage}[t]{0.33\linewidth}
			\centering
			\includegraphics[width=2.0in]{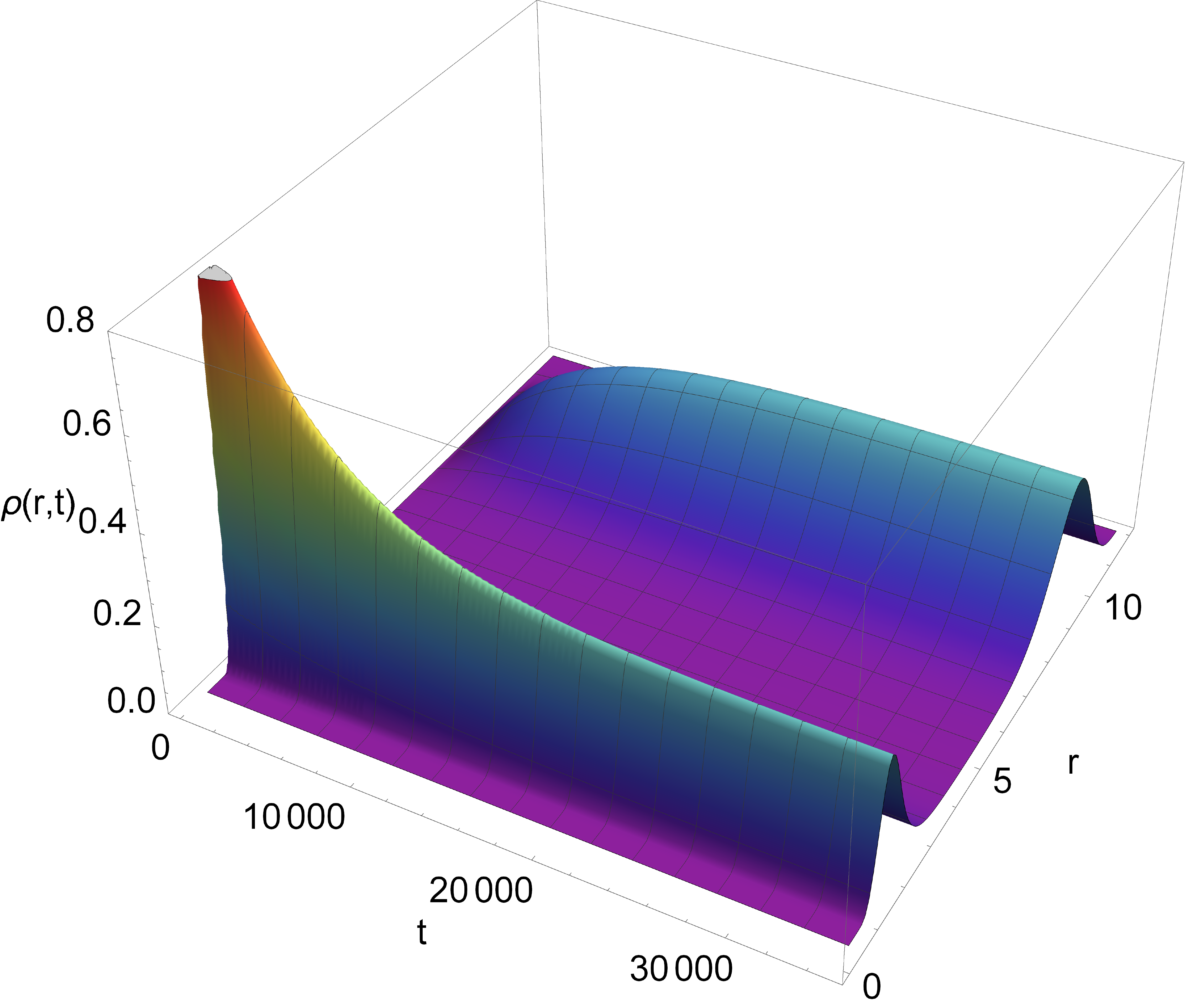}
		\end{minipage}
	}%
	\quad
	\subfigure[$\eta=0,~T_{E}=0.029806$\label{ptr4}]{
		\begin{minipage}[t]{0.32\linewidth}
			\centering
			\includegraphics[width=2.0in]{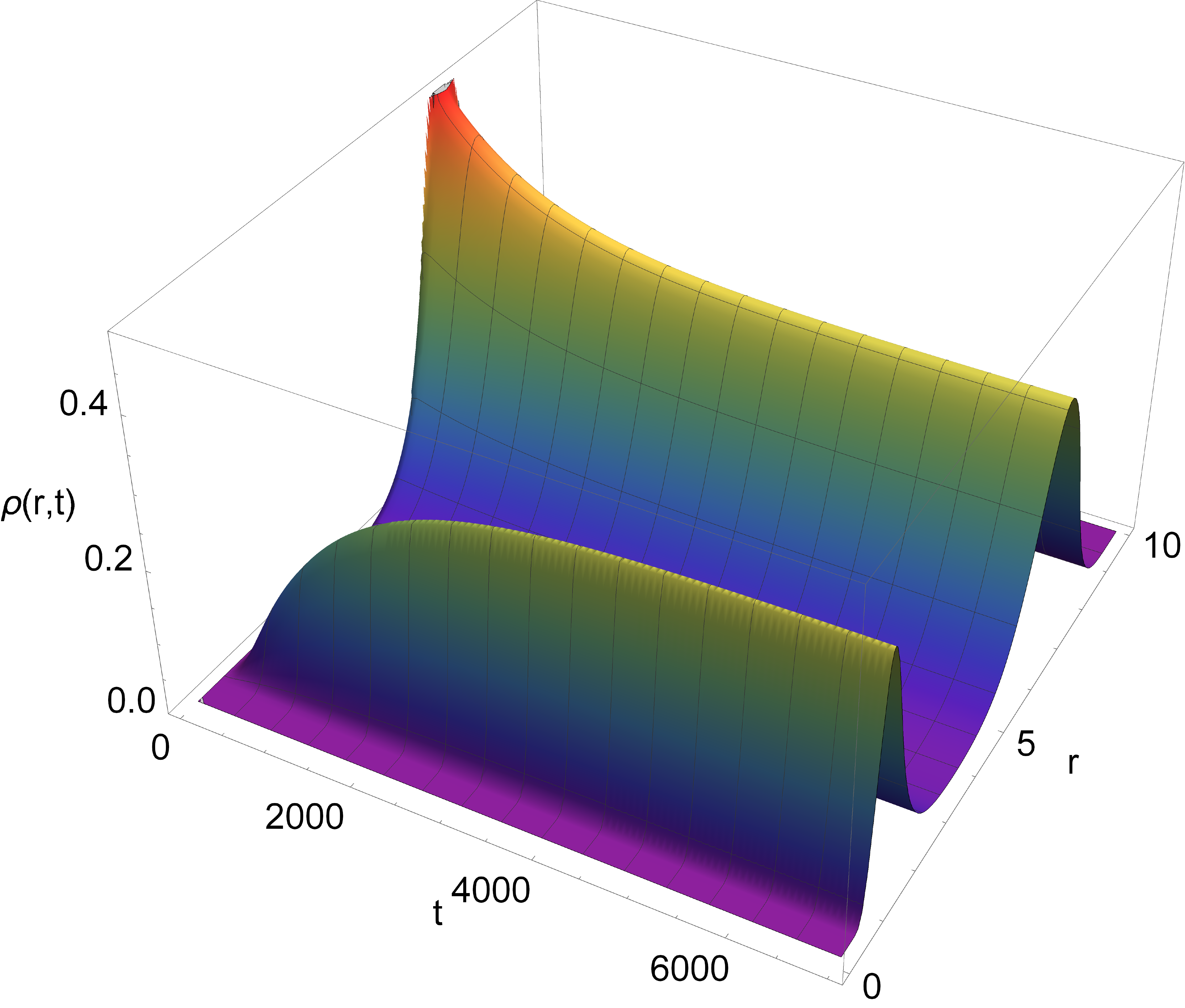}
		\end{minipage}
	}%
    \subfigure[$\eta=0.3,~T_{E}=0.027124$\label{ptr5}]{
    	\begin{minipage}[t]{0.32\linewidth}
    		\centering
    		\includegraphics[width=2.0in]{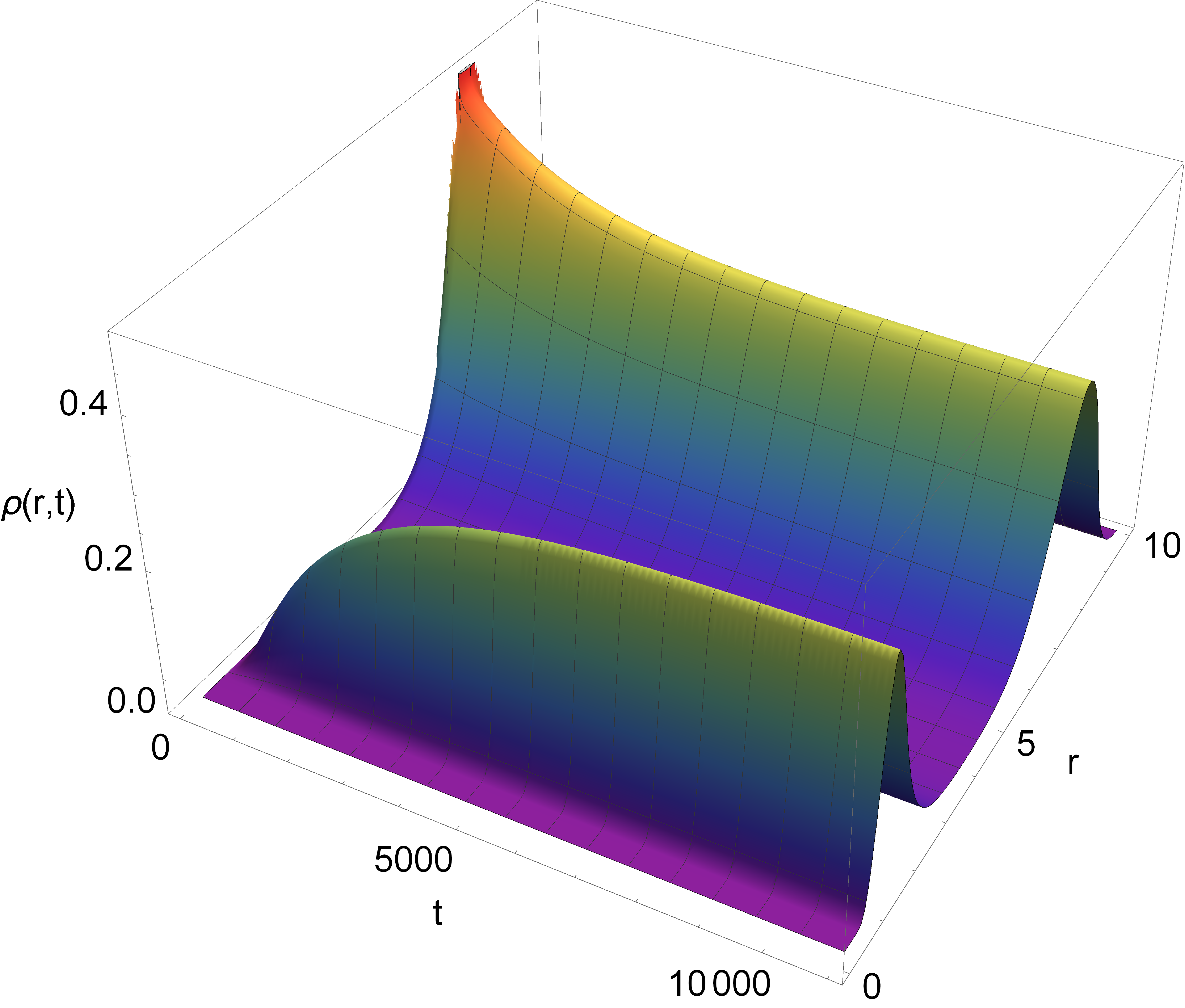}
    	\end{minipage}
    }%
    \subfigure[$\eta=0.5,~T_{E}=0.022355$\label{ptr6}]{
    	\begin{minipage}[t]{0.32\linewidth}
    		\centering
    		\includegraphics[width=2.0in]{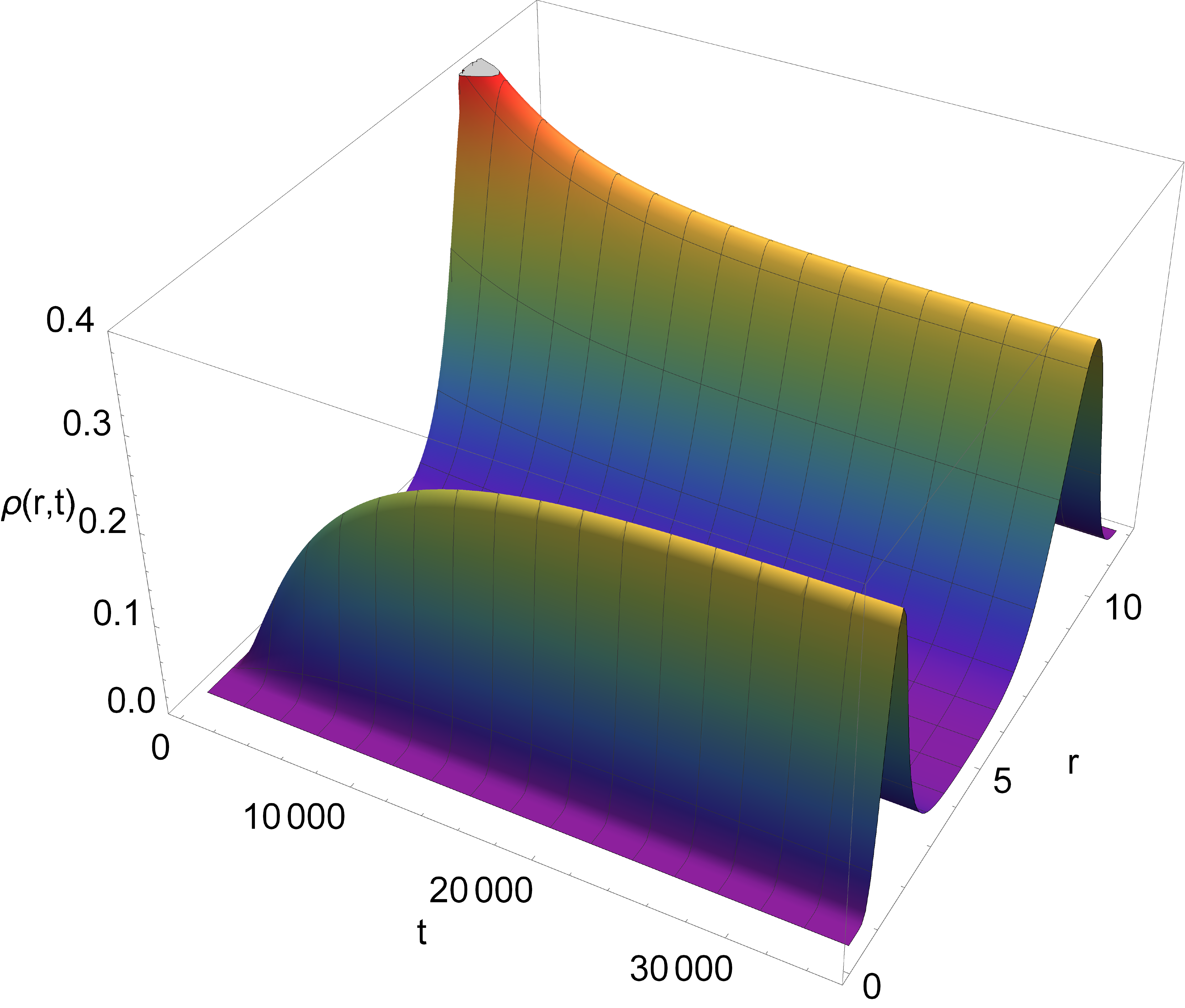}
    	\end{minipage}
    }%
	
	\caption{Time evolution of the probability distribution $\rho\left(r, t\right)$ for the charged AdS BH with a global monopole. For the up panels, the Gaussian wave pocket is located at the SBH state. For the down panels, the Gaussian wave pocket is located at the LBH state.}
	\label{fig:ptr1}
\end{figure}

\begin{figure*}
	\centering

\subfigure[$\eta=0,~T_{E}=0.029806$]{
	\begin{minipage}[t]{0.33\linewidth}
		\centering
		\includegraphics[width=1.97in]{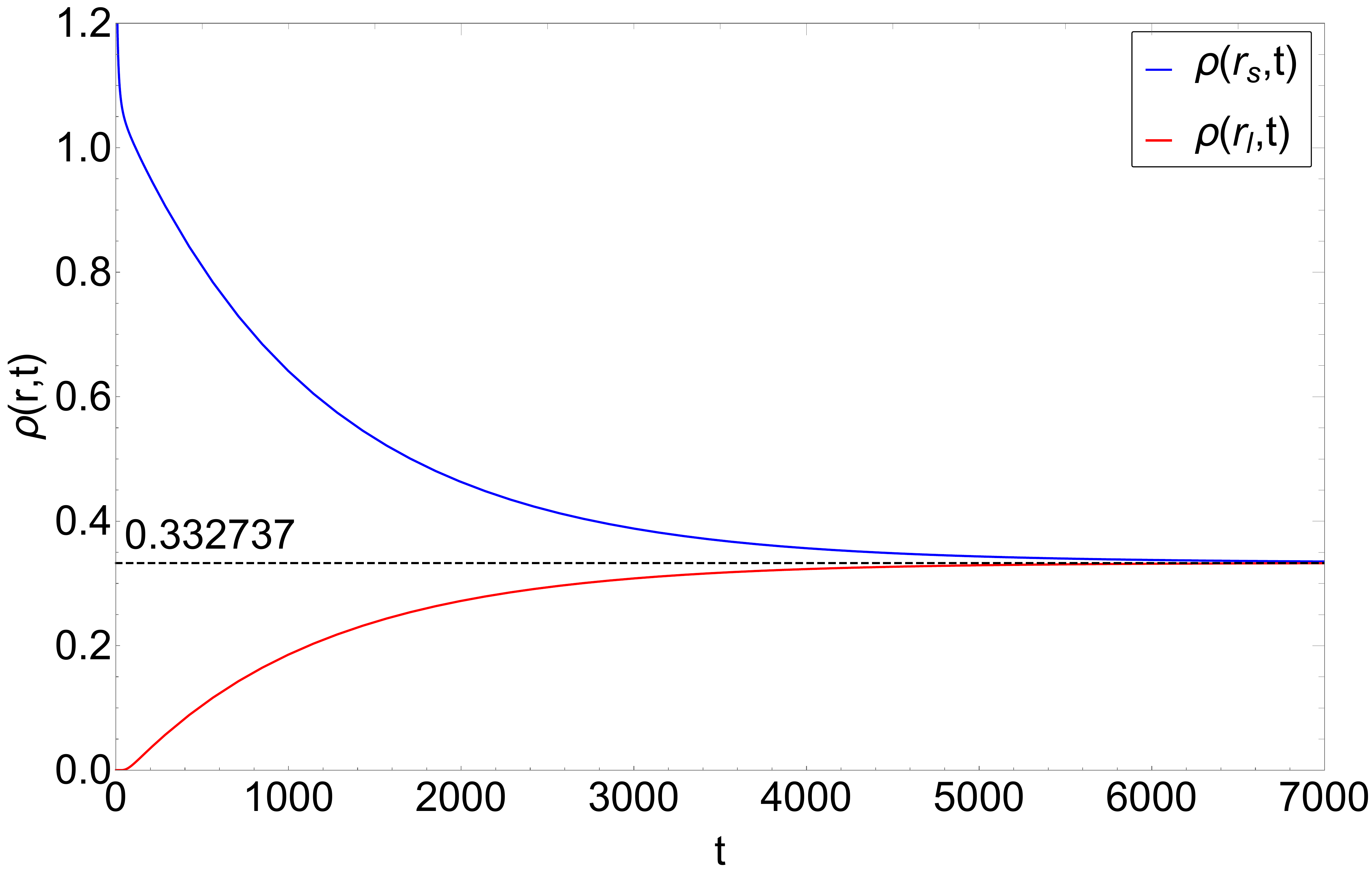}
	\end{minipage}\label{4a}
}%
\subfigure[$\eta=0.3,~T_{E}=0.027124$]{
	\begin{minipage}[t]{0.33\linewidth}
		\centering
		\includegraphics[width=1.92in]{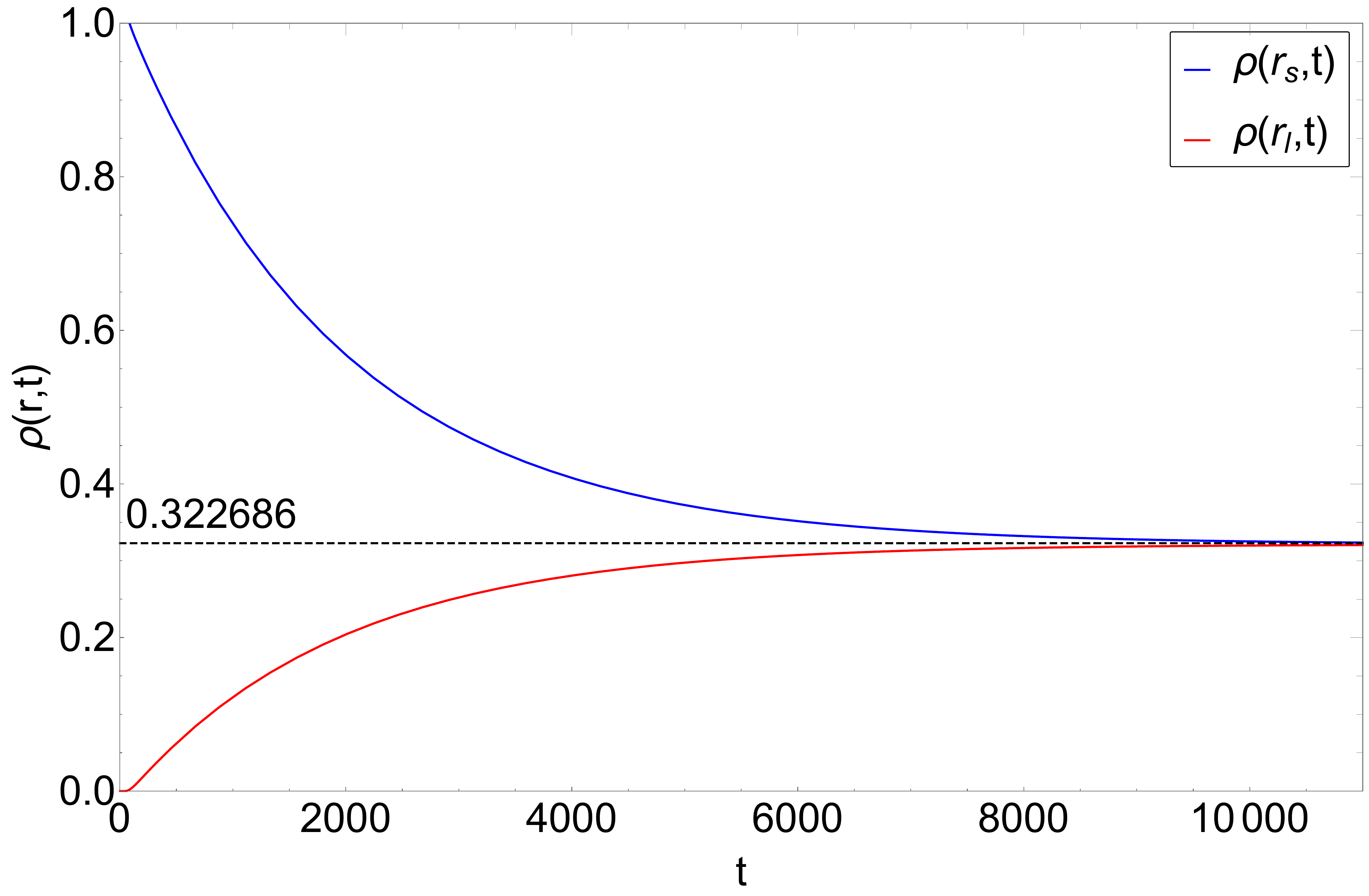}
	\end{minipage}\label{4b}
}%
\subfigure[$\eta=0.5,~T_{E}=0.022355$]{
	\begin{minipage}[t]{0.33\linewidth}
		\centering
		\includegraphics[width=2.0in]{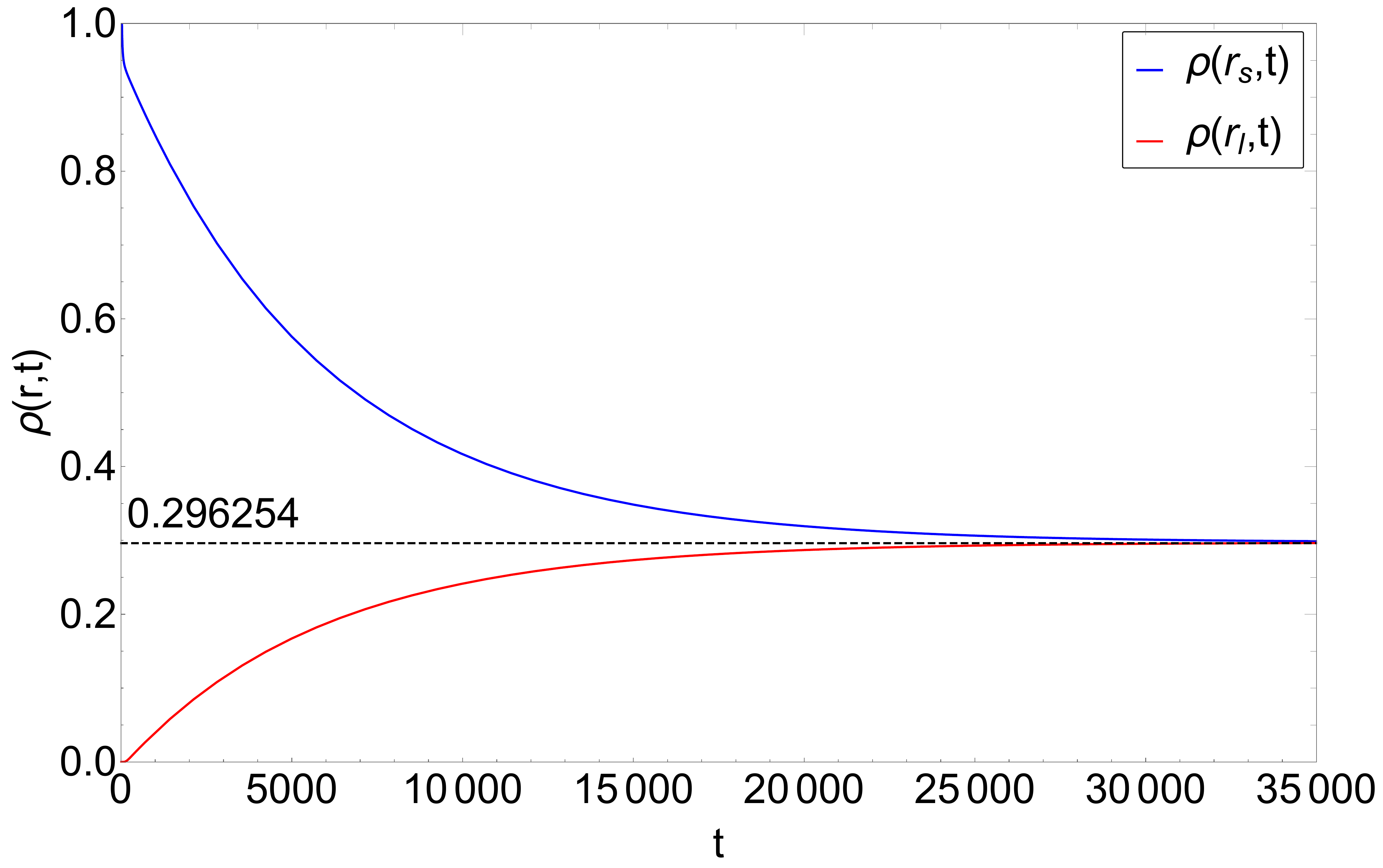}
	\end{minipage}\label{4c}
}%
\quad
\subfigure[$\eta=0,~T_{E}=0.029806$]{
	\begin{minipage}[t]{0.33\linewidth}
		\centering
		\includegraphics[width=1.97in]{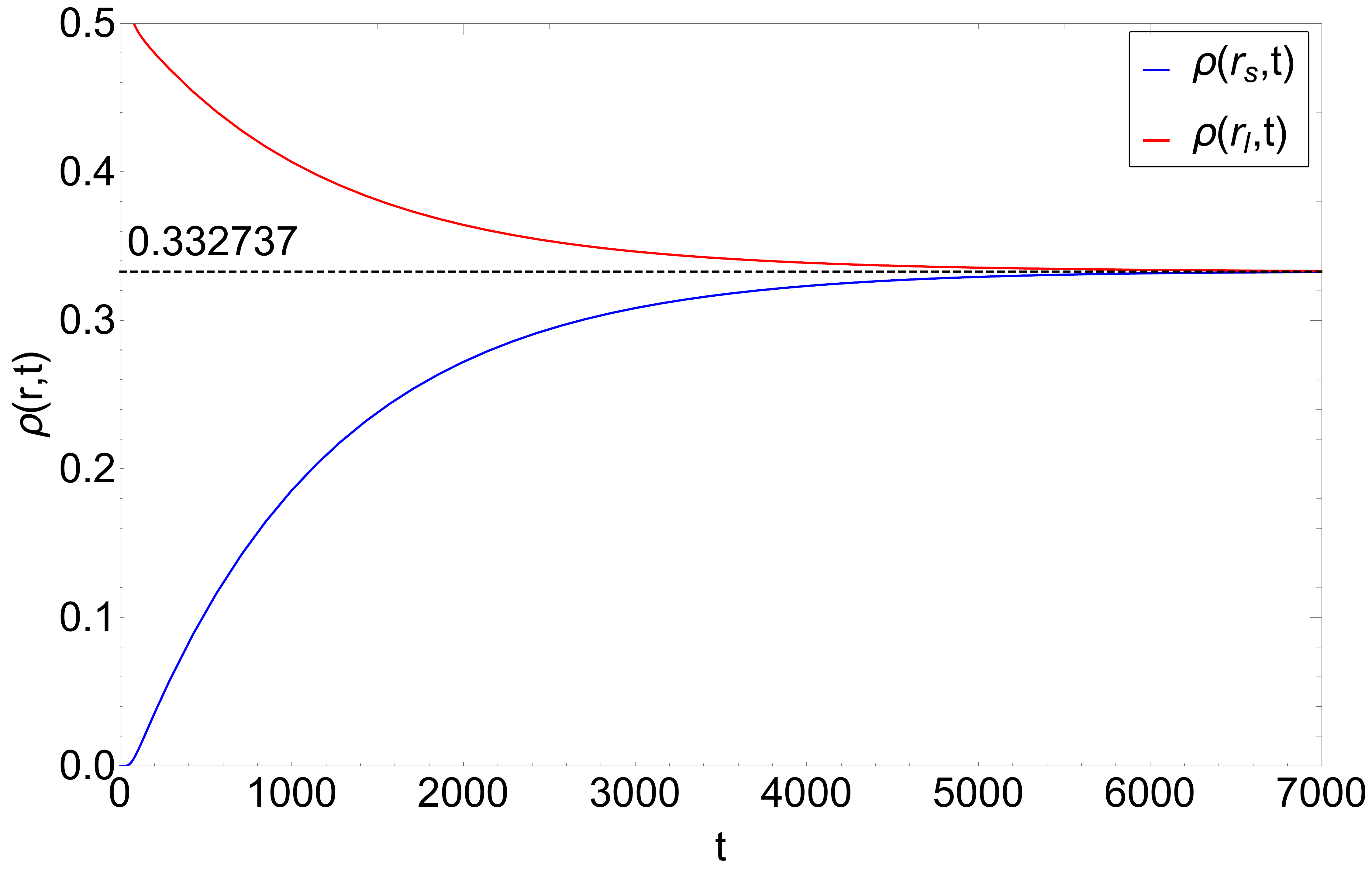}
	\end{minipage}\label{4d}
}%
\subfigure[$\eta=0.3,~T_{E}=0.027124$]{
	\begin{minipage}[t]{0.33\linewidth}
		\centering
		\includegraphics[width=1.92in]{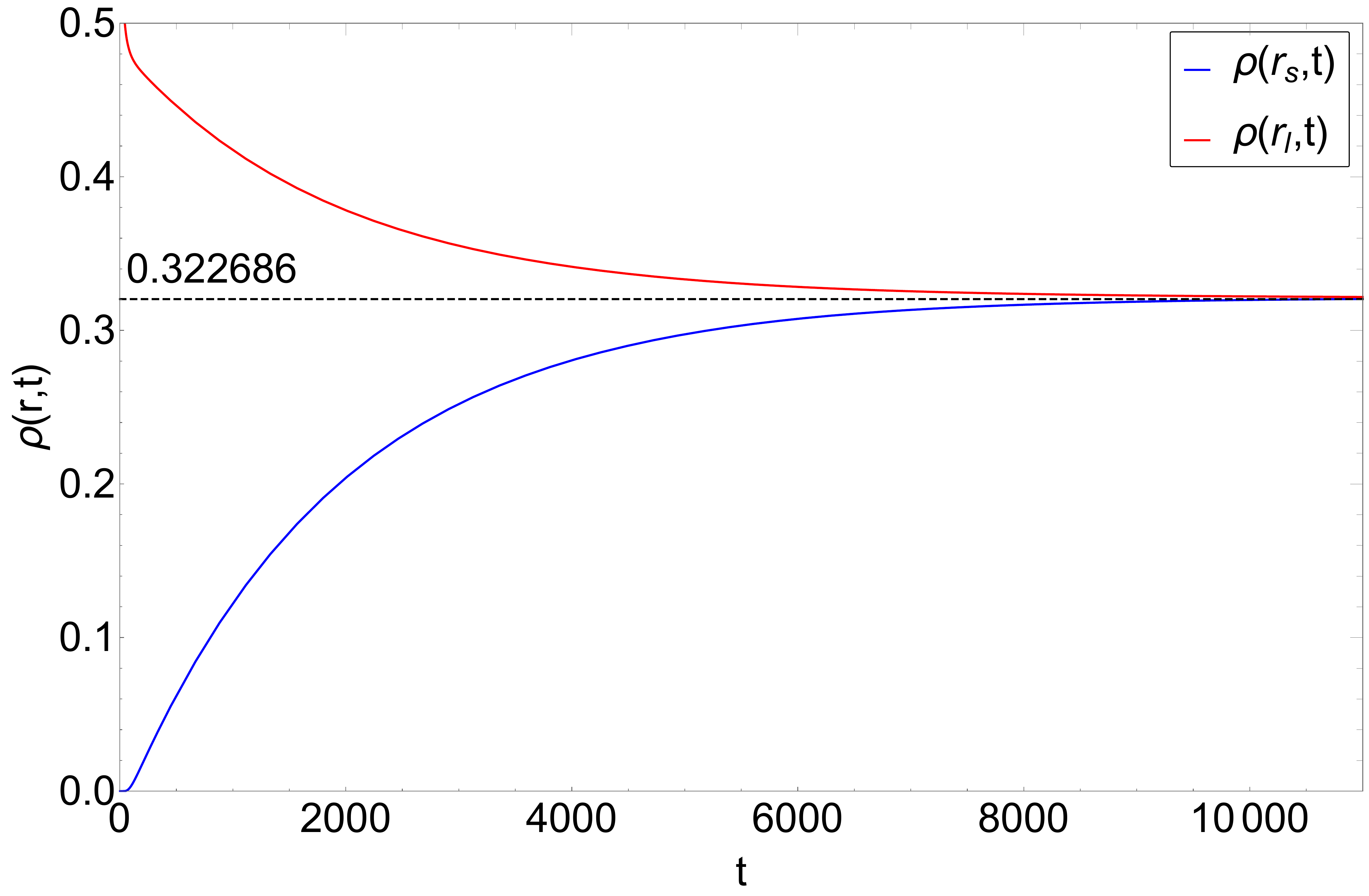}
	\end{minipage}\label{4e}
}%
\subfigure[$\eta=0.5,~T_{E}=0.022355$]{
	\begin{minipage}[t]{0.33\linewidth}
		\centering
		\includegraphics[width=2.0in]{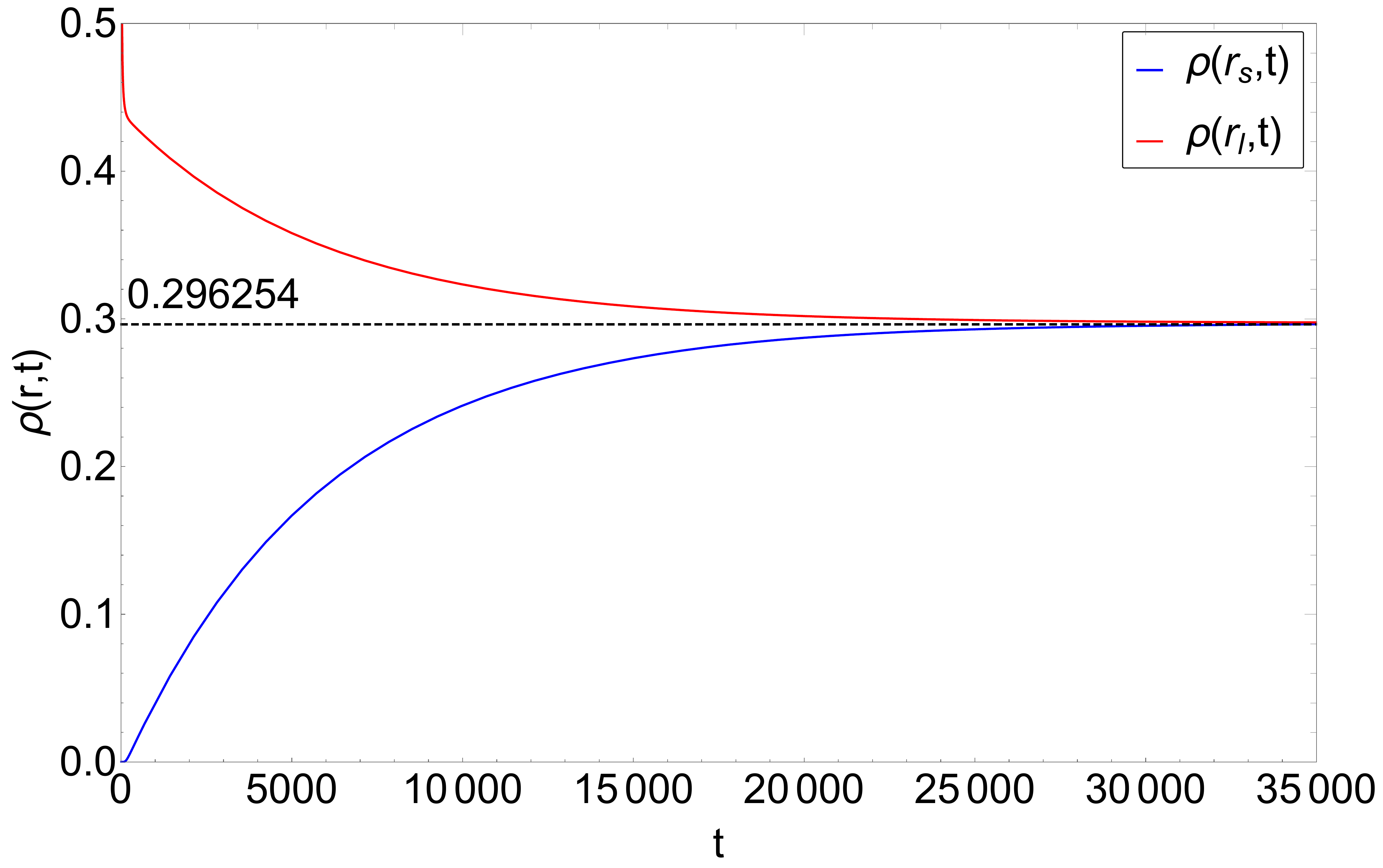}
	\end{minipage}\label{4f}
}%
  
	\centering
	\caption{Time evolution of the probability distribution $\rho\left(r, t\right)$ for given SBH and LBH radii. For the up panels, the Gaussian wave pocket is located at the SBH state. For the down panels, the Gaussian wave pocket is located at the LBH state.}
	\label{fig:phot} 
\end{figure*}

In Fig.~\ref{fig:ptr1}, we plot the probability distribution diagrams of the charged AdS BH with a global monopole via numerical calculation. Note that in order to avoid numerical instability we have replaced the boundaries at $r=0$ and $r=\infty$ with $r=0.1$ and $r=12$, respectively. From Figs.~\ref{ptr1}, ~\ref{ptr2} and~\ref{ptr3}, one can find when the probability distribution evovles from the SBH state, the probability distribution of the SBH state decreases gradually, while the probability distribution of the LBH state increases from $\rho\left(r_{l}, t\right)=0$. In Figs.~\ref{ptr4}, ~\ref{ptr5} and~\ref{ptr6}, we can observe a similar phenomenon for the case that the probability distribution evovles from the LBH state. The more intuitive descriptions of the time evolution of the probability distribution for the SBH and LBH  states are shown in Fig.~\ref{fig:phot}. Obviously, no matter what the value of the monopole parameter $\eta$ is, the system will always reach a steady-state distribution after a period of time and the probability distributions of the SBH and LBH states are equal. However, acoording to Fig.~\ref{fig:phot}, it is also clear that the global monopole has effects on the thermodynamic PT of BHs. Comparing Figs.~\ref{4a},~\ref{4b} and~\ref{4c}, one can find the initial SBH state with a global monopole (such as $\eta=0.5$ and $\eta=0.3$) evolves more slowly than the one without a global monopole ($\eta=0$). For the initial LBH state with a global monopole, one can obtain the same conlusion (see Figs.~\ref{4d},~\ref{4e} and~\ref{4f}). Besides that, as the monopole parameter $\eta$ increases, it takes longer for the system to reach equilibrium.

\subsection{First passage process}\label{sec4}
In the BH PT, the process that the initial SBH (LBH) state reaches the IBH state for the first time is called the first passage process, and the time it takes is called the first passage time. In order to study the first passage process, we need to solve the Fokker-Planck equation not only with the reflecting boundary condition but also with the absorbing boundary condition (i.e., $\rho\left(r_{i}, t\right)=0$), where $r_{i}$ is the IBH radius.

\begin{figure}[h]
	\centering

	\subfigure[$\eta=0,~T_{E}=0.029806$\label{fptr1}]{
		\begin{minipage}[t]{0.33\linewidth}
			\centering
			\includegraphics[width=2.0in]{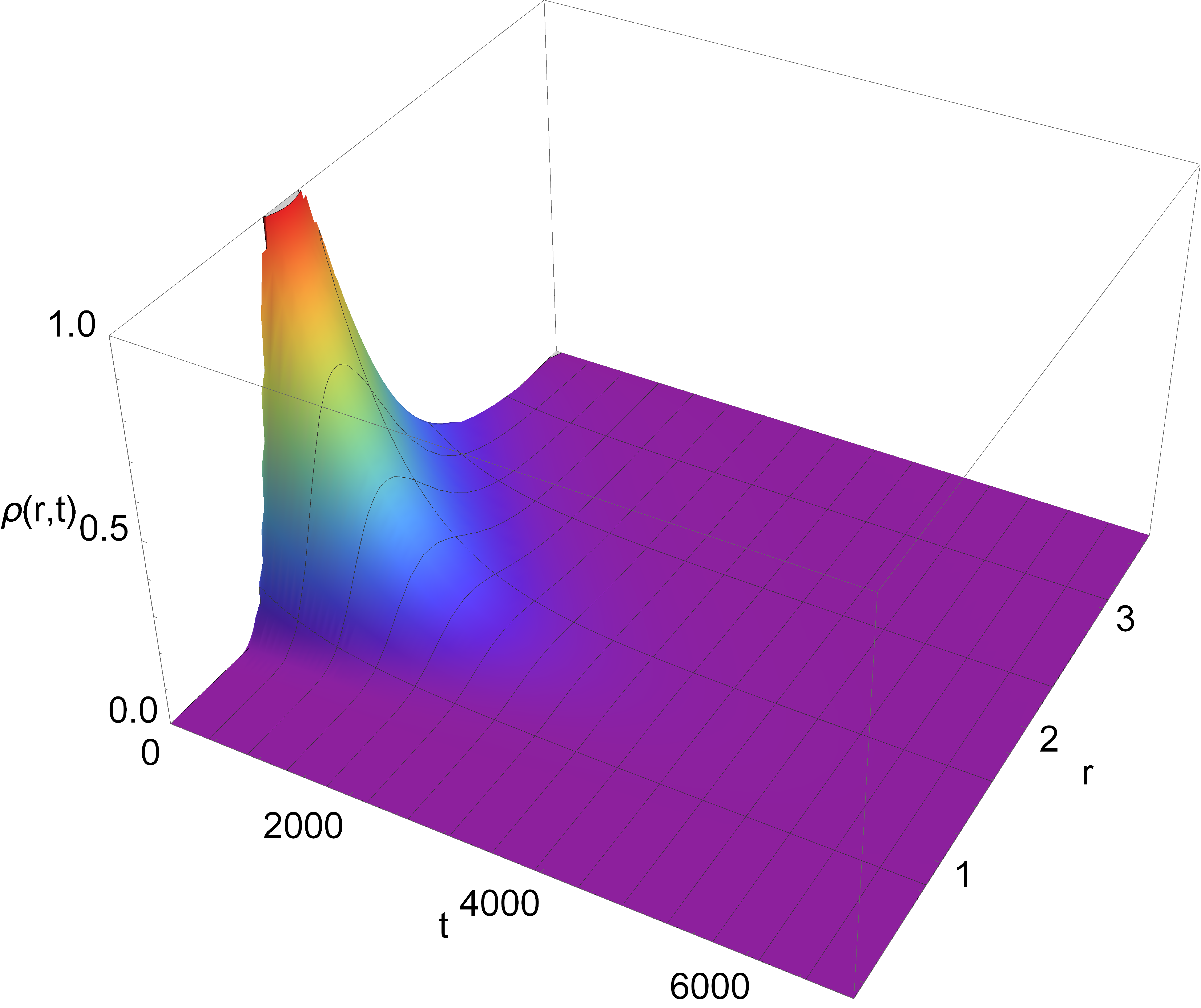}
		\end{minipage}\label{5a}
	}%
	\subfigure[$\eta=0.3,~T_{E}=0.027124$\label{fptr2}]{
		\begin{minipage}[t]{0.33\linewidth}
			\centering
			\includegraphics[width=2.0in]{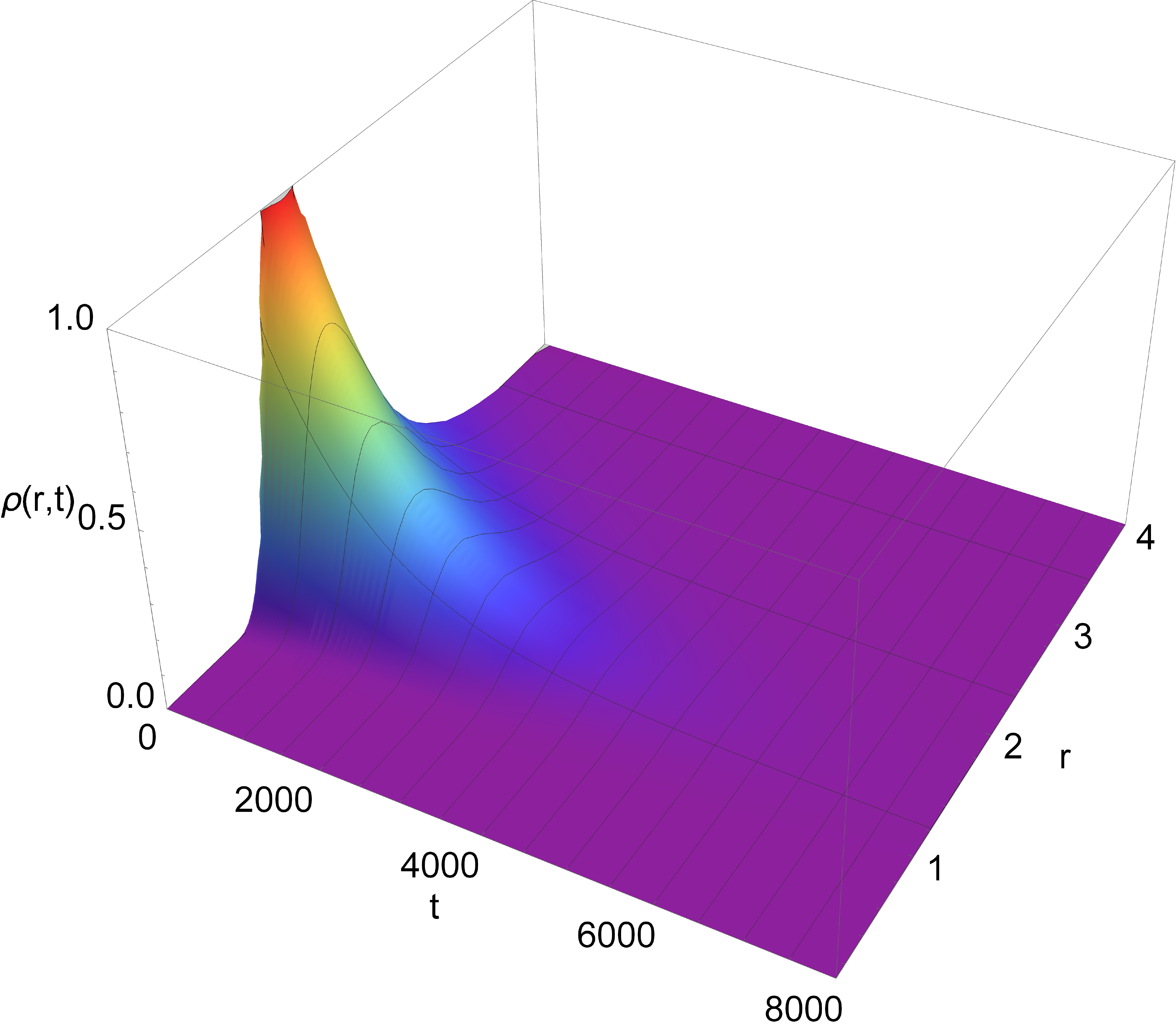}
		\end{minipage}\label{5b}
	}%
\subfigure[$\eta=0.5,~T_{E}=0.022355$\label{fptr3}]{
	\begin{minipage}[t]{0.33\linewidth}
		\centering
		\includegraphics[width=2.0in]{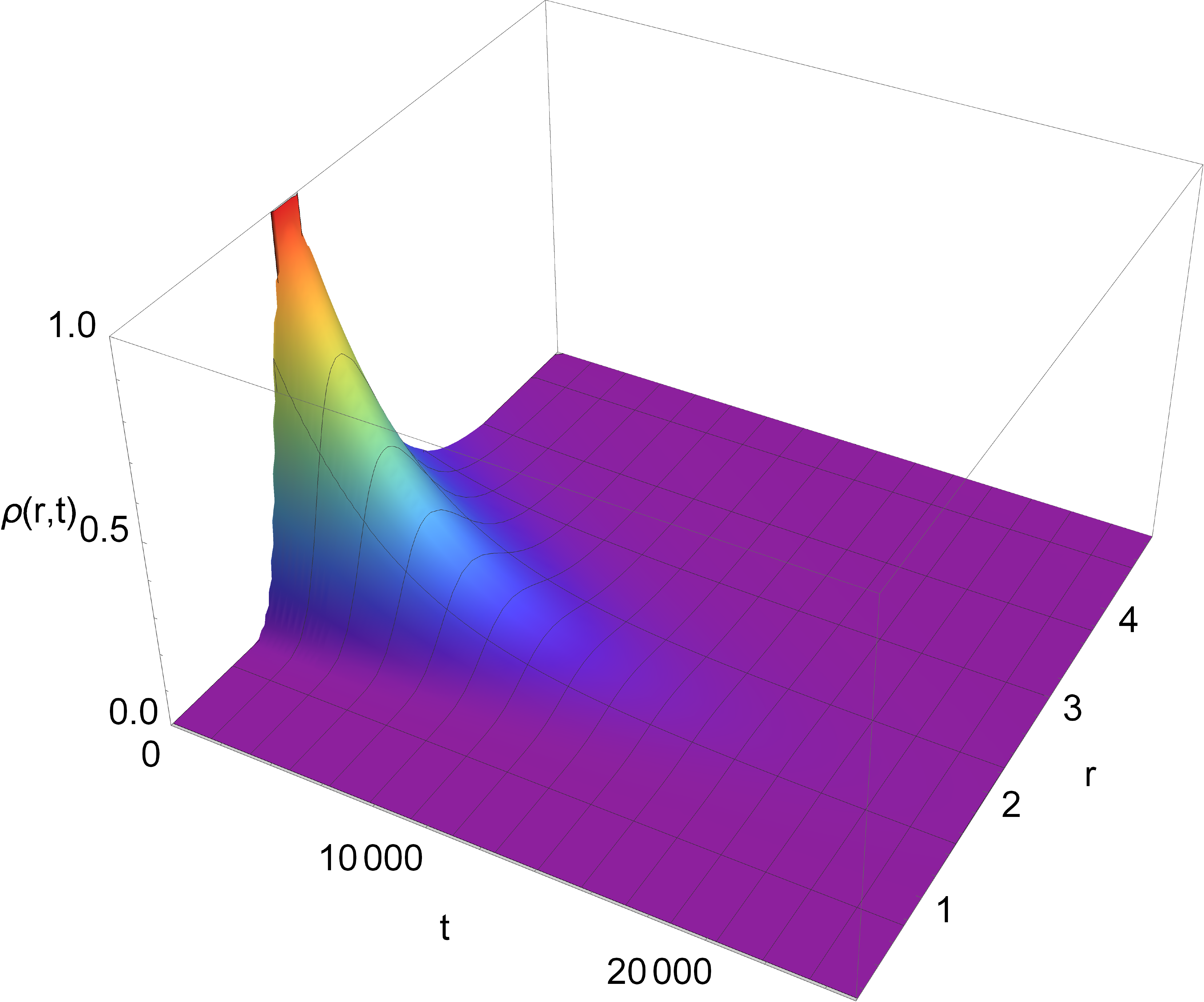}
	\end{minipage}\label{5c}}
	\quad
	\subfigure[$\eta=0,~T_{E}=0.029806$\label{fptr4}]{
		\begin{minipage}[t]{0.33\linewidth}
			\centering
			\includegraphics[width=2.0in]{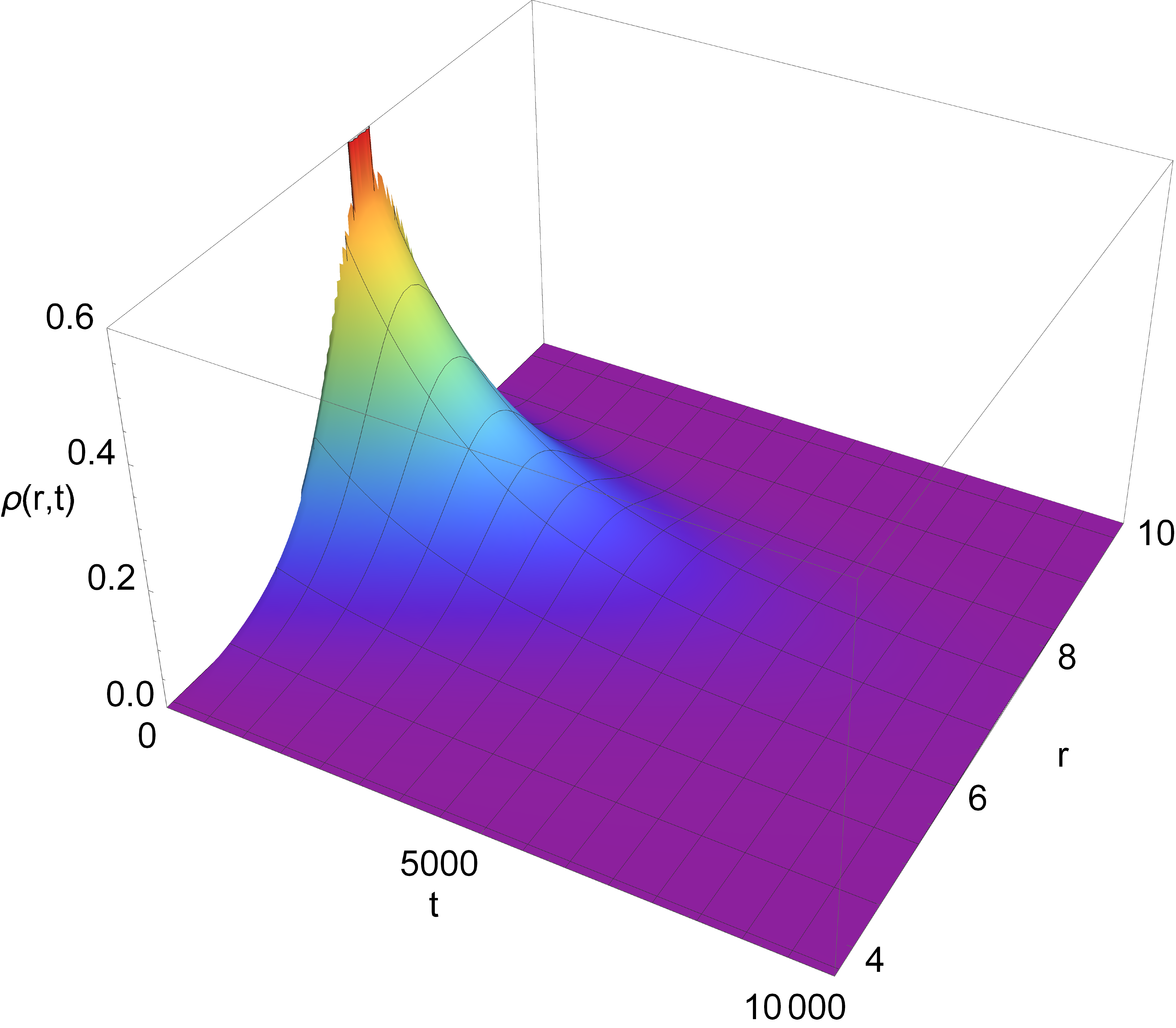}
		\end{minipage}
	}%
	\subfigure[$\eta=0.3,~T_{E}=0.027124$\label{fptr5}]{
		\begin{minipage}[t]{0.33\linewidth}
			\centering
			\includegraphics[width=2.0in]{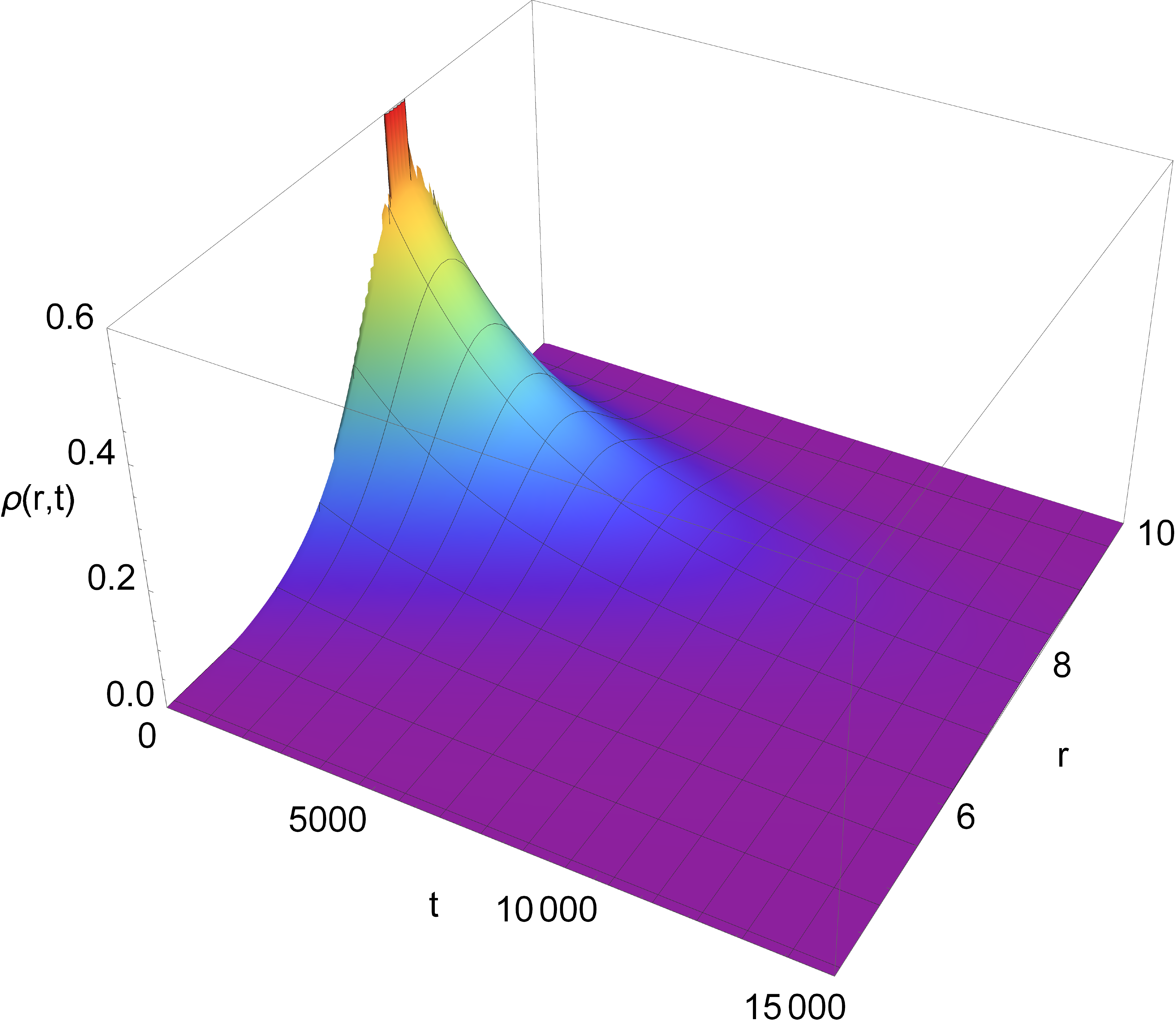}
		\end{minipage}
	}%
	\subfigure[$\eta=0.5,~T_{E}=0.022355$\label{fptr6}]{
		\begin{minipage}[t]{0.33\linewidth}
			\centering
			\includegraphics[width=2.0in]{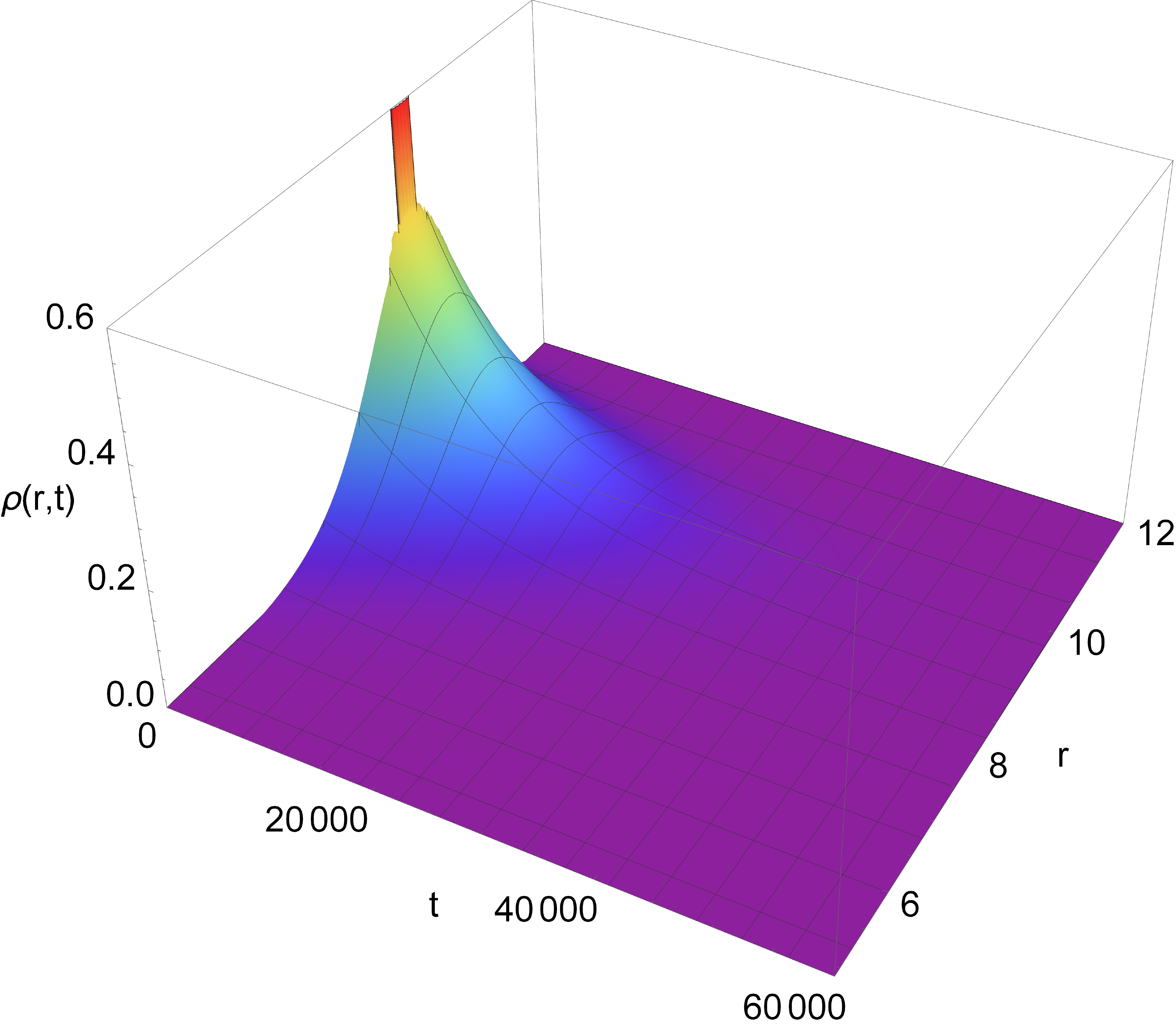}
		\end{minipage}\label{5d}}
	
	\caption{Time evolution of the probability distribution $\rho(r, t)$ of the first passage process for the charged AdS BH with a global monopole. For the up panels, the Gaussian wave pocket is located at the SBH state. For the down panels, the Gaussian wave pocket is located at the LBH state.}
	\label{fptr}
\end{figure}

Taking the reflecting and absorbing boundary conditions into the Fokker-Planck equation, we can solve the probability distribution $\rho(r_{s/l}, t)$ numerically. In Fig.~\ref{fptr}, we plot the time evolution of the probability distribution of the first passage process for the charged AdS BH with a global monopole. It is clear from Fig.~\ref{fptr} that the probability distribution of both the SBH and LBH states decays quickly. Moreover, by comparing Figs.~\ref{5a}, \ref{5b}, and \ref{5c}, we can conclude that the probability distribution of the SBH state with a global monopole evolves more slowly than the one without a global monopole. And as the monopole parameter $\eta$ increases, it also takes longer for the system to reach equilibrium. These affects of the global monopole on the the probability distribution of the BH state are consistent with the ones in Sec.~\ref{secfp} where we only consider the reflecting boundary condition.

\begin{figure*}
	\centering
	
	\subfigure[]{
		\begin{minipage}[t]{0.5\linewidth}
			\centering
			\includegraphics[width=2.5in]{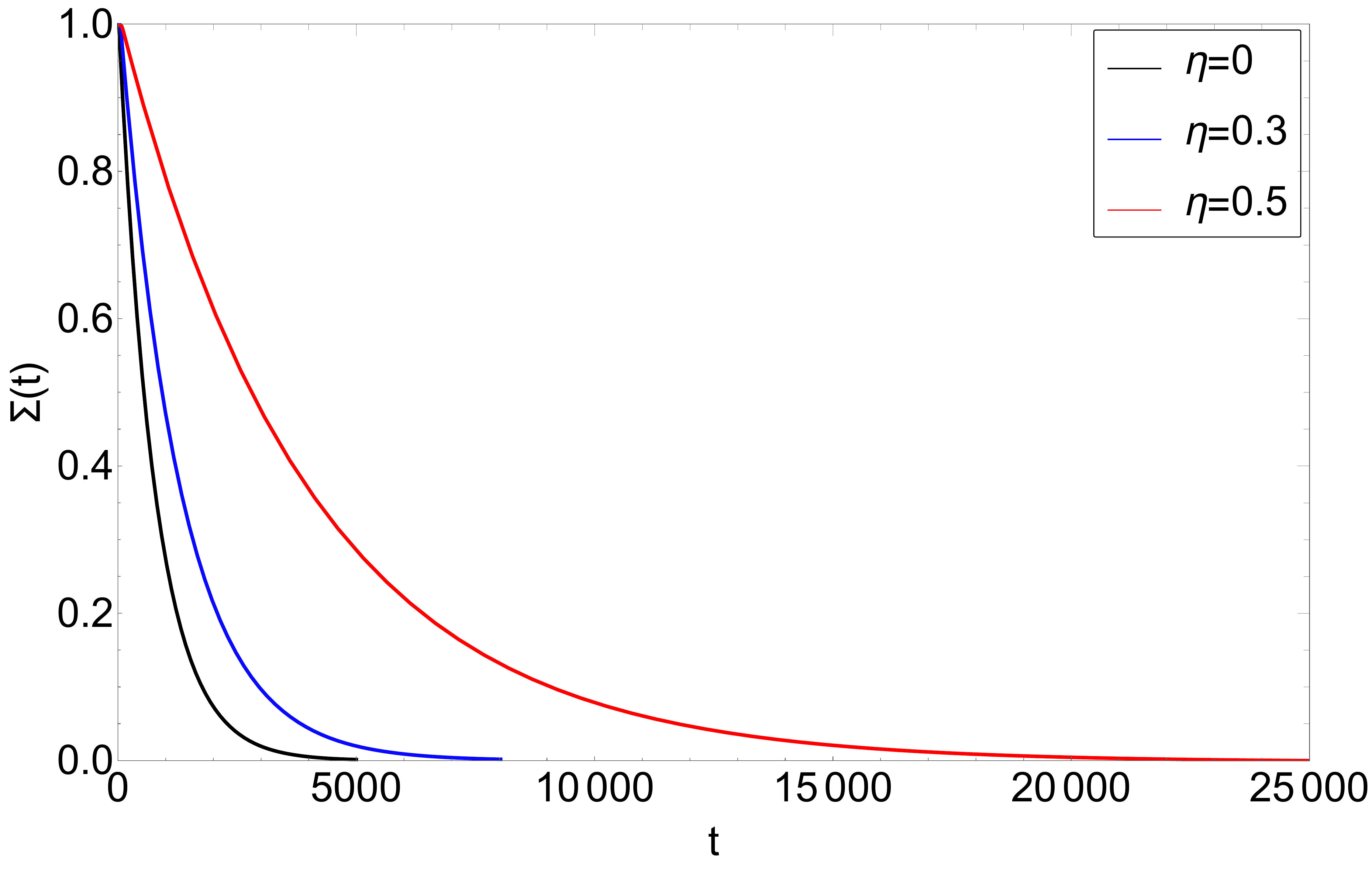}
		\end{minipage}%
	}%
	\subfigure[]{
		\begin{minipage}[t]{0.5\linewidth}
			\centering
			\includegraphics[width=2.5in]{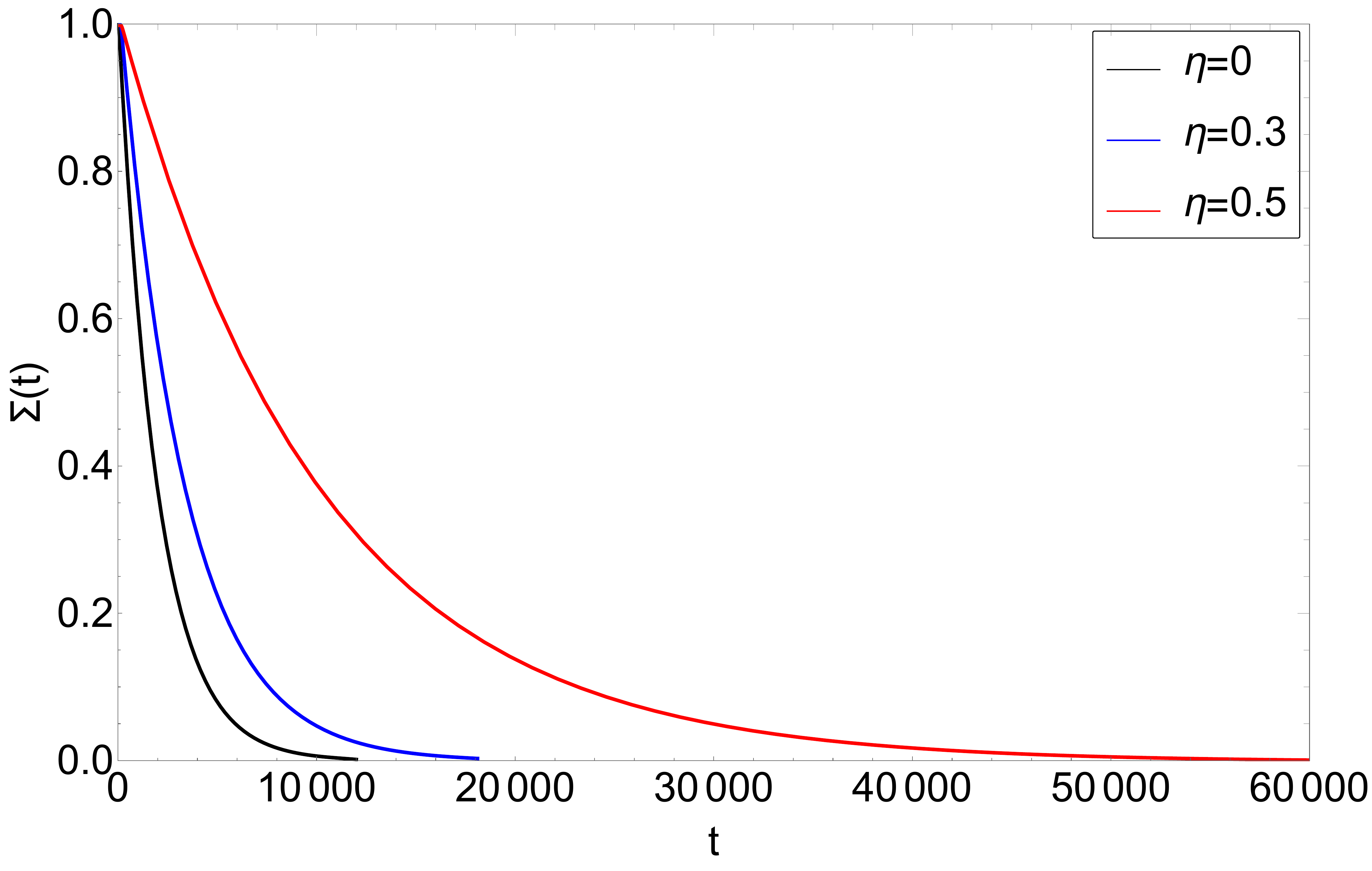}
		\end{minipage}%
	}%

	\centering
	\caption{Time evolution of the probability $\Sigma(t)$. In the figure (a), the Gaussian wave pocket is located at the SBH state, and the figure (b) denotes the LBH state.}
	\label{fig:sigma} 
\end{figure*}

Now, we define the probability that the present state of the BH is still in the system at time $t$ as follows:
\begin{equation}
	\Sigma(t)=\int_{0}^{r_{m}} \rho(r, t) d r,
\end{equation}
which also can be interpreted that the BH at time $t$ has not made the first passage process. For the charged AdS BH with a global monopole, we plot the $\Sigma(t)-t$ diagrams in Fig.~\ref{fig:sigma}. One can see that $\Sigma(t)$ will vanish  after a period of time, which means that the BH has made the first passage process. It is clear that in the presence of the global monopole, the time evolution of the probability $\Sigma(t)$  is longer. As the global monopole parameter increases, the effect is more prominent.

\begin{figure*}
	\centering
	
	\subfigure[]{
		\begin{minipage}[t]{0.5\linewidth}
			\centering
			\includegraphics[width=2.5in]{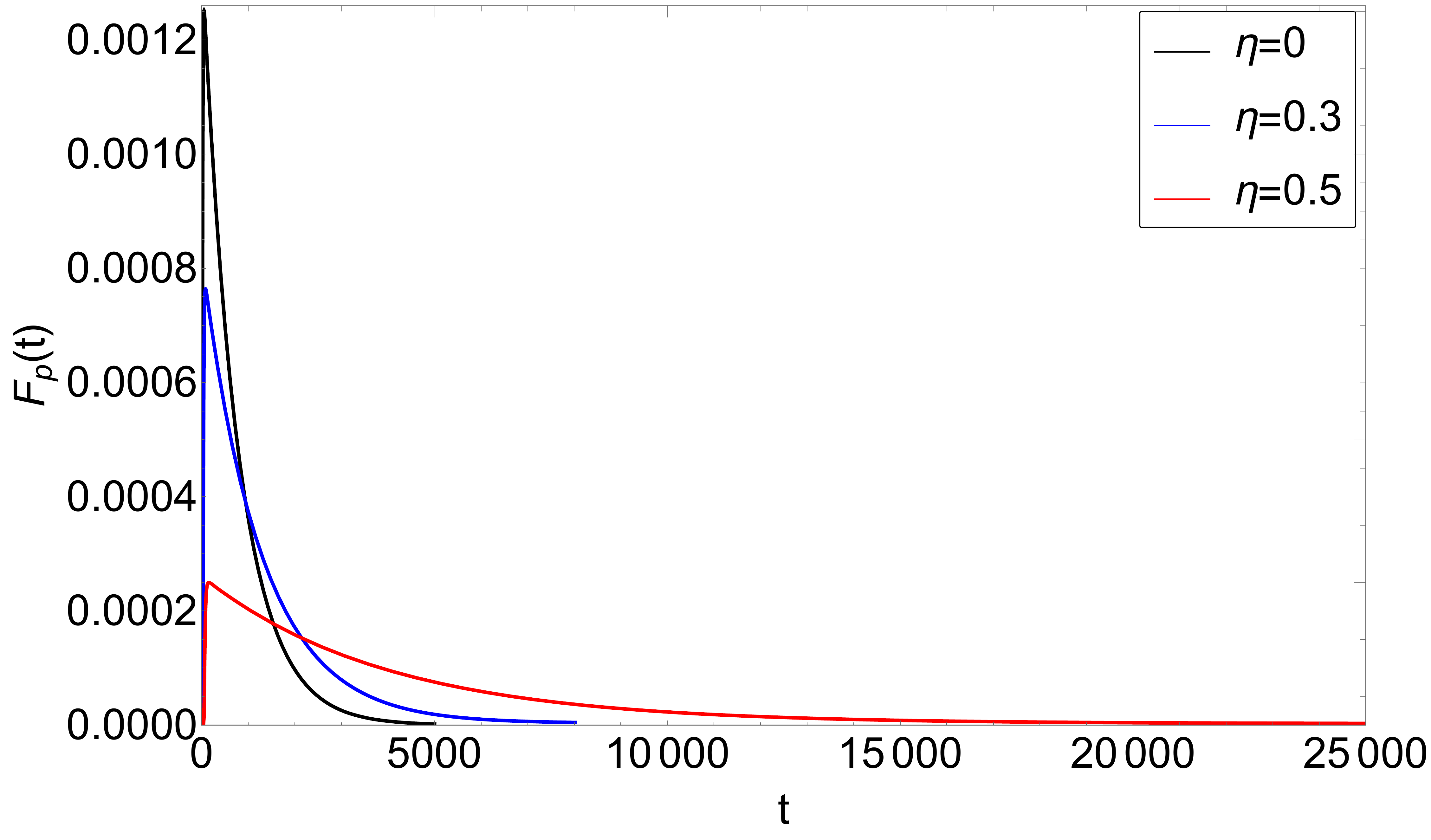}
		\end{minipage}%
	}%
	\subfigure[]{
		\begin{minipage}[t]{0.5\linewidth}
			\centering
			\includegraphics[width=2.5in]{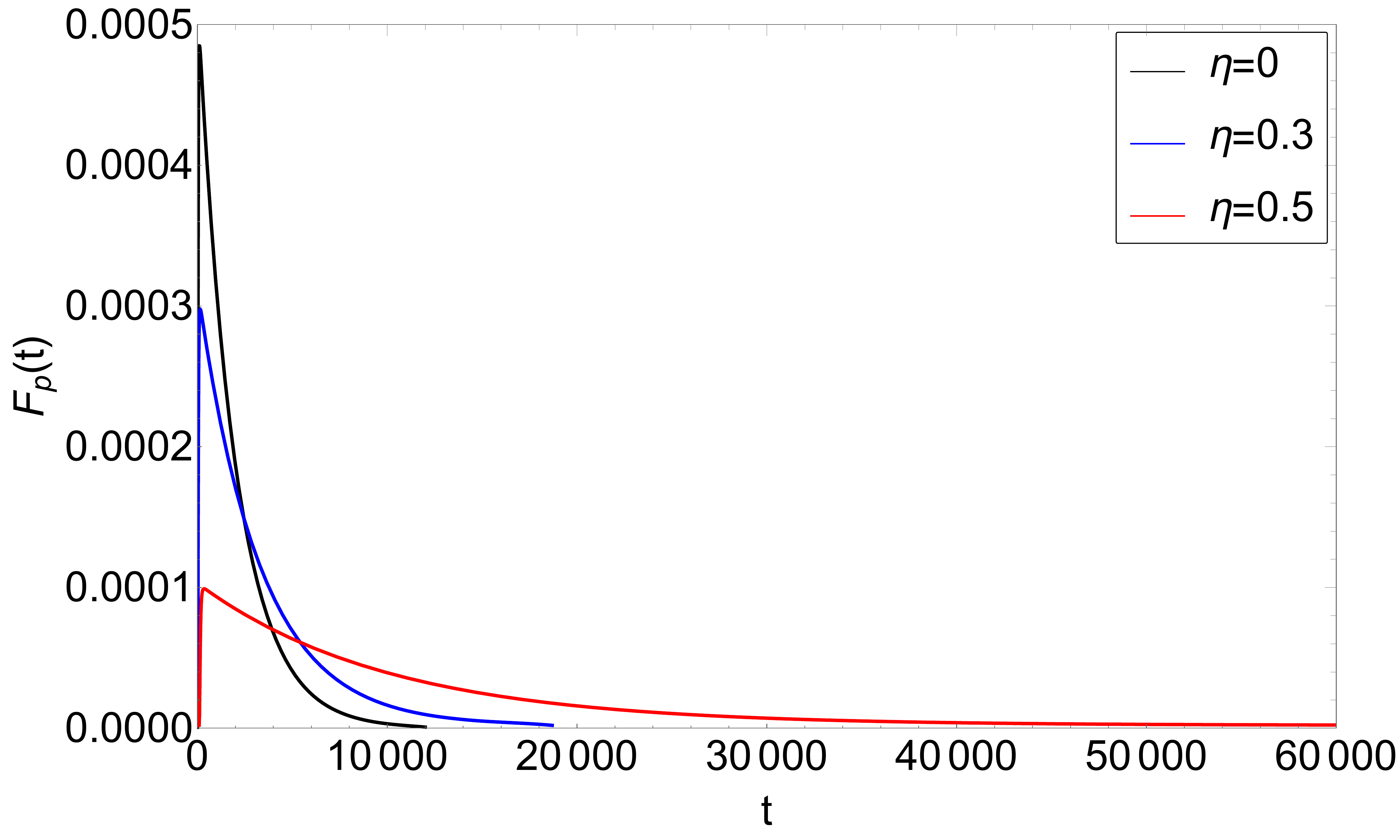}
		\end{minipage}%
	}%

	\centering
	\caption{First passage time $F_{p}(t)$. In the figure (a), the Gaussian wave pocket is located at the SBH state, and the figure (b) denotes the LBH state.}
	\label{fig:Fpt} 
\end{figure*}

Moreover, the distribution of the first passage time can be defined as~\cite{Li:2020nsy}
\begin{equation}
F_{\mathrm{p}}(t)=-\left.D \frac{\partial}{\partial r_{h}} \rho(r_{h}, t)\right|_{r=r_{m}}.
\end{equation}
We plot the $F_{p}(t)-t$ diagrams in Fig.~\ref{fig:Fpt}. It can be found that for each $\eta$ there is a single peak near the beginning time, which indicates that most of the first passage events are completed in a short time. It is easy to that the impact of the global monopole on the first passage time is similar to the previous cases.

To study further the effects of the global monopole on the thermodynamic PT of the BH, we calculate the mean first passage time, which is defined as~\cite{Li:2020nsy}
\begin{equation}
\langle t\rangle=\int_{0}^{\infty} t F_{p}(t) d t.
\end{equation}
The values of the mean first passage time for different monopole parameters are listed in Tabel.~\ref{tab1}. We have calculated seven sets of the values of the mean first passage time and the reuslts show that the mean first passage time increases with the monopole parameter.

\begin{table}[]
	\centering
	\caption{The mean first passage time for different monopole parameters}
	\begin{tabular}{|l|l|l|l|l|l|l|l|}
		\hline
		$\eta$                                                                  & 0    & 0.1  & 0.2  & 0.3  & 0.4  & 0.5  & 0.6   \\ \hline
		$\left.\langle t\rangle\right|_{\mathrm{SBH}\rightarrow \mathrm{LBH}} $ & 774  & 845  & 958  & 1278 & 1937 & 3962 & 11290 \\ \hline
		$\left.\langle t\rangle\right|_{\mathrm{LBH}\rightarrow \mathrm{SBH}} $ & 1922 & 2086 & 2429 & 3108 & 5053 & 9933 & 24857 \\ \hline
	\end{tabular}
	\label{tab1}
\end{table}
 

\section{Conclusions} \label{sec5}
This paper aims to research the effects of a global monopole on the thermodynamic PT of the charged AdS BH. First, we briefly reviewed the thermodynamics of the charged AdS BH with a global monopole. Then, we studied the Gibbs free energy of the BH and the dynamical properties of the thermodynamic PH of the BH in the context of the Gibbs free energy landscape. It is found that the AdS BH with a global monopole has similar thermodynamic and dynamical characteristics to ordinary AdS BHs~\cite{Li:2020khm,Li:2020nsy,Yang:2021ljn,Wei:2021bwy,Li:2021vdp,Lan:2021crt,Wei:2020rcd}.

The influence of the global monopole on the BH can be analyzed by solving the Fokker-Planck equation. Therefore, we first obtained the solution of the Fokker-Planck equation with the reflecting boundary condition. The probability distribution diagrams signify that the system will eventually reach a steady state, at which the probability distributions of the SBH and LBH states are equal. Interestingly, it is found that the global monopole could slow down the decay rate of the initial BH state. And the larger the monopole parameter is, the more obvious the effect will be. In order to investigate further the impact of the global monopole on the thermodynamic PT of the AdS BH, then we studied the first passage process of the dynamical PT of the BH. We resolved the Fokker-Planck equation with the reflecting and absorbing boundary conditions. The time evolution of the probability distribution with the monopole parameter varying shows that the probability distribution always decays rapidly, but the decay rate decreases with the increase of the monopole parameter. Finally, we calculated the mean first passage time for different monopole parameters. The result shows that the mean first passage time monotonically increases as the monopole parameter increases, which means that the effect of the  monopole parameter on the mean first passage time is similar to the effect on the probability distribution. 

For the charged AdS BH with a global monopole, since the monopole parameter has significant influence on the mean first passage time, if we can figure out the observational effect of the mean first passage time of the thermodynamic PT in the future, we may further constrain the monopole parameter in related astronomical observations. Recently, there are some new theoretical and observational progresses on the relationship between the BH thermodynamics and the BH shadow~\cite{EventHorizonTelescope:2022xnr,EventHorizonTelescope:2022vjs,EventHorizonTelescope:2022wok,EventHorizonTelescope:2022exc,EventHorizonTelescope:2022urf,EventHorizonTelescope:2022xqj,Tsukamoto:2014tja,Wei:2017mwc,Kumar:2018ple,Zhang:2019glo,Allahyari:2019jqz,Belhaj:2020nqy,Wang:2021vbn,Cai:2021uov,EventHorizonTelescope:2021dqv,Guo:2022yjc}. The thermodynamic PT of the charged AdS BH with a global monopole might be reflected in the shadow radius, which can give an upper limit to the monopole parameter of the BH.

\acknowledgments
This work was supported by the National Natural Science Foundation of China (Grant Nos. 11873001 and 12047564), the Fundamental Research Funds for the Central Universities under Grand No.2021CDJZYJH-003, and the Postdoctoral Science Foundation of Chongqing (Grant No. cstc2021jcyj-bsh0124).


\appendix

\end{CJK}
\end{document}